\documentclass[reprint, amsmath,amssymb,aps,prx]{revtex4-2}
\usepackage{physics}
\usepackage{enumitem}
\usepackage{graphicx}% Include figure files
\usepackage{dcolumn}
\usepackage{bm}
\usepackage[dvipsnames]{xcolor}
\usepackage{hyperref}
\usepackage{cleveref}
\usepackage{color}
\usepackage{amsmath}
\usepackage{tikz-cd}
\usepackage{caption,float}
\graphicspath{{figures/}}

\newcommand\myshade{85}
\colorlet{mylinkcolor}{violet}
\colorlet{mycitecolor}{YellowOrange}
\colorlet{myurlcolor}{Aquamarine}

\hypersetup{
  linkcolor  = mylinkcolor!\myshade!black,
  citecolor  = mycitecolor!\myshade!black,
  urlcolor   = myurlcolor!\myshade!black,
  colorlinks = true,
}

\newtheorem{definition}{Definition}[section]
\newtheorem{theorem}{Theorem}[section]

\begin{document}

\title{Generalized quantum similarity learning}

\author{\href{http://www.santoshkumarradha.me}{Santosh Kumar Radha}}
\email{santosh@agnostiq.ai}
\author{Casey Jao}%
\email{casey@agnostiq.ai}

\affiliation{%
 Agnostiq Inc. 325 Front St W, Toronto, ON M5V 2Y1\\
}%

\date{\today}

\begin{abstract}

The similarity between objects is significant in a broad range of areas. While
it can be measured using off-the-shelf distance functions, they may fail to
capture the inherent meaning of similarity, which tends to depend on the
underlying data and task. Moreover, conventional distance functions limit the
space of similarity measures to be symmetric and do not directly allow
comparing objects from different spaces. We propose using quantum networks (\textit{GQSim}) for
learning task-dependent (\textit{a})symmetric
similarity between data that need not have
the same dimensionality. We analyze the properties of such similarity function
analytically (for a simple case) and numerically (for a complex case) and show
that these similarity measures can extract salient features of the data. We also demonstrate that the similarity measure derived using this technique is $(\epsilon,\gamma,\tau)$-good, resulting in theoretically guaranteed performance. Finally, we conclude by applying this technique for three relevant applications - Classification, Graph Completion, Generative modeling.
\end{abstract}

\maketitle

\section{Introduction}

Notions of (dis)similarity are fundamental to learning.
For example, they are
implicit in probabilistic models that use dissimilarity
computations to derive model parameters. In contrast,
$k-$nearest neighbor (KNN) or Support Vector Machines (SVM) methods explicitly
find training instances similar to the input. Notions of similarity also play a
fundamental role in human learning. It is well known that people can perceive
different degrees of similarity. The semantics of such judgments may depend on
both the task at hand and the particulars of data. Since manually tuning such
similarity functions can be difficult and tedious for real-world problems, the
notion of automatically \textit{learning} task/data-specific similarity
($\mathcal{S}$) from labeled data have been introduced
\cite{schultz2004learning,shalev2004online,chechik2009online,bellet2013survey,nicolae2015joint}.
 In general, these learning methods are based on the intuition that a
good similarity function should assign a large/small score to a pair of points
in the same/different classes.

The most common way to model similarity is to
use a \emph{distance metric} $d$ as a model for the
desired similarity measure. By definition, $d$ obeys the properties
$d(x,x)=0$,
$d(x_1,x_2) = d(x_2, x_1) \geq0$, $d(x_1,x_2)+d(x_2,x_3)\geq d(x_1,x_3)$, where
$x \in
\mathcal{X}$ with $\mathcal{X}$ the data space. An important effect of this is
the transitive nature of the similarity function \textit{i.e.} $\mathcal{S}$
derived from $d$ obeys the following, if $\mathcal{S}(x_1,x_2)=1$ and
$\mathcal{S}(x_2,x_3)=1$, then $\mathcal{S}(x_1,x_3)=1$. The class of
similarity maps attainable by these methods is just a subset of general
similarity maps.

Another major technique used in classical
machine learning is manifold-learning. This method aims to learn a
low-dimensional structure in data under the assumption that the data lie on a
(possibly non-linear) manifold. Examples of such methods
include\cite{bengio2003out} local linear embedding\cite{roweis2000nonlinear},
multidimensional scaling\cite{cox2008multidimensional}, Laplacian
Eigenmaps\cite{belkin2003laplacian} etc. These methods, by nature, produce only
symmetric similarity measures \textit{i.e.}
$\mathcal{S}(x_1,x_2)=\mathcal{S}(x_2,x_1)$.

\begin{figure}[h]
  \includegraphics[width=0.7\linewidth]{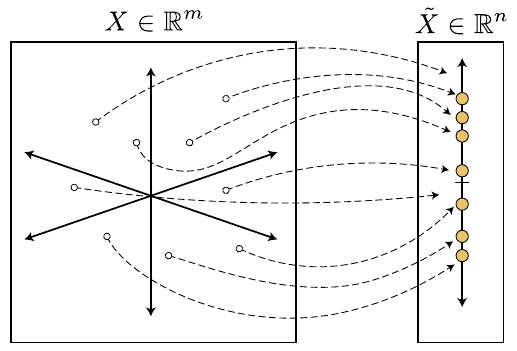}
  \caption{Learning relations between
  data in different spaces}
  \label{fig:hero}
\end{figure}

Meanwhile, much work has been done to marry quantum computing and classical
machine learning techniques to construct practical quantum algorithms
compatible with the Noisy Intermediate-Scale Quantum (NISQ) era. One approach
is to use parametrized quantum circuits (PQC) to define a hypothesized class of
functions that potentially might offer access to a class of models beyond what
is possible with classical models. Such quantum learning/variational algorithm have been used for various applications\cite{kerenidis2019quantum,dallaire2018quantum,radha2021quantum,coyle2021quantum,radha2021quantumwick}. Theoretical works have been successful in
showing such quantum models exist for specific synthetic
problems\cite{liu2021rigorous,sweke2021quantum,huang2021power}. Specific
classes of such quantum machine learning (QML) algorithms can be mathematically
connected to their classical counterpart -  kernel
methods\cite{schuld2021quantum,pronobis2020kernel}. In these methods, the
learning is based on the kernels that are calculated between two given data
points after mapping them to a latent space. This kernel can be made of PQCs,
which, when trained, define distances on the latent
space\cite{lloyd2020quantum}.

In this paper, we show that a general framework of such quantum embedding maps
expands the class of algorithms from \textit{distance metric} learning to
(a)symmetric-similarity learning ones, which we call \textit{GQSim}. This general class of QML algorithms has
striking similarities with the classical similarity learning methods, including
Siamese networks\cite{bromley1993signature}. We show that these extensions
allow for more general properties like intransitivity, asymmetry, etc.,
enabling us to learn a larger class of similarity maps than just distance-based
similarity. Finally, one advantage of this technique is the ability
to compare objects belonging to possibly \emph{different spaces}, as shown
schematically in
Figure~\ref{fig:hero}. Formally, given
$\tilde{x}\in \mathbb{R}^{\tilde{n}}$ and $x\in \mathbb{R}^{n}$ data instances
from two different sources/base space, and set of triplets
$(\tilde{x}_i,x_i,y_i)$ where $y_i:=\mathcal{S}(\tilde{x}_i,x_i)$, the task is
to
predict $y_j$ for a pair of unseen instances $(\tilde{x}_j,x_j)$ with certain
non trivial properties for $\mathcal{S}$ (like asymmetry). Note that even
though we have $y_i\in \mathbb{R}_{+}$, the same technique can easily be
extended to non-regression type problems by having $y_i\in \{0,1\}$. Such
learning applications abound in nature; for instance, learning the sentence
similarity between two different languages would generally be asymmetric and
involve two different spaces.

We start by formulating the problem at hand and its underlying theory in
\autoref{sec:formalism}. In \autoref{sec:discussion}, we first discuss
analytically various details of the \textit{GQSim} methods,
including the effect of partial measurement. We illustrate this by solving a
toy-problem analytically. Second, in \autoref{subsec:ND} we pick a
numerical problem of learning similarity between two different subspaces made
up of synthetic images and points in $\mathbb{R}^2$ to demonstrate the
generalizability of the learned model. \autoref{sec:app} discusses similarity
learning in various applied settings, and finally, we summarize our findings in
\autoref{sec:summery}.

\section{Formalism}\label{sec:formalism}
For sets $X$ and $\tilde{X}$, assume we are given set of elements $A \subset X
\times \tilde{X}$ and the corresponding set of similarity measure $Y$ where
$Y:=\mathcal{S}(x,\tilde{x})$ for all $(x,\tilde{x})\in A $ with $Y \in
\mathbb{R}_{+}$. Our goal is to learn the model $\mathcal{S}$ that is used to
predict the similarity between unseen elements in $X \times \tilde{X}$. We
first start by generalizing metric learning for multi-subspace setting after
which we introduce the framework for learning asymmetric similarity. Even
though $\mathcal{S}\in \mathbb{R}_{+}$, for the sake of discussion, we will
restrict ourselves to $\mathcal{S}$ being binary values of either similar or
dissimilar. Learning of $\mathcal{S}$ is done by using parameterized quantum
circuits, details of which are discussed in \autoref{subsubsec:exp setup}.

\begin{figure}[h!]
  \includegraphics[width=\linewidth]{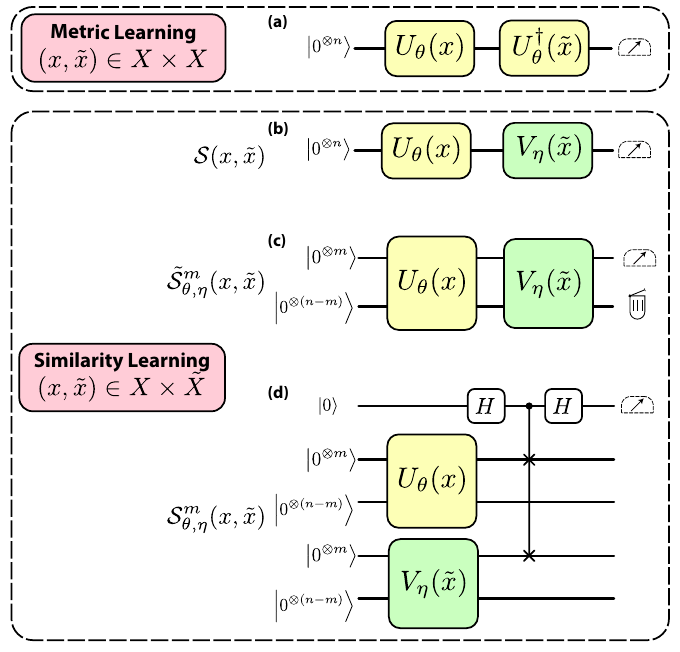}
  \caption{Various similarity learning techniques are discussed in this paper.
  \textit{(a)} Metric learning where the resulting function forms a distance
  metric. \textit{(b)} Multi-space similarity learning with different
  embeddings, \textit{(c/d)} Multi-space similarity learning gives us access to
  asymmetric functions (Please refer to section~\ref{sec:gsimlearn} for more
  details)}\label{fig:all circuits}
\end{figure}

\subsection{Multi-subspace metric learning}\label{sec:multimetric}
Given $\mathcal{E}_l, \mathcal{D}_k \subset X \times \tilde{X}$  with cardinality $l,k$ such that pairs $(x_i,\tilde{x}_i) \in \mathcal{E}_l$ are deemed \emph{related} ($y_i=1$) and $(x_i, \tilde{x}_i) \in \mathcal{D}_k$ are \emph{unrelated} ($y_i=0$). Then we try to learn a distance function \[d: X \times \tilde{X} \to \mathbb{R}_+\] such that
\begin{align}
  \label{e:learning_criterion}
  \begin{cases}
    d(x,\tilde{x}) \le \delta & (x,\tilde{x}) \in \mathcal{E}_l\\
    d(x,\tilde{x}) \ge \varepsilon^{-1} & (x,\tilde{x}) \in \mathcal{D}_k
  \end{cases}
\end{align}
for some small constants $\delta, \varepsilon> 0$.
Here, $d$ need not satisfy the usual axioms of a distance
metric, such as symmetry, which may not even make sense
as the inputs come from different spaces. The problem can also be phrased in
terms of a \emph{similarity measure} $\mathcal{S}$, which simply reverses the inequalities: $\mathcal{S}(x,\tilde{x})$ should be large when $x$ and $\tilde{x}$ are similar and small when they are dissimilar.

We consider $d$ of the form
\begin{align}
  d(x,\tilde{x}) := \| f_\theta (x) - g_\eta (\tilde{x})\|_\mathcal{H}.
\end{align}
where the \emph{feature maps} $f_\theta: X \to \mathcal{H}$, $g_\eta: \tilde{X}
\to \mathcal{H}$ encode the raw input data in a common Hilbert space
$\mathcal{H}$. Concretely, let $H = (\mathbb{C}^2)^{\otimes n}$ be the Hilbert space of a quantum computer with $n$ qubits, and let $\mathcal{H} = \mathcal{L}(H)$ be the space of complex-linear operators on $H$ equipped with the Hilbert-Schmidt inner product. Following \cite{havlivcek2019supervised}, define the feature maps
\begin{align}
  \label{e:feature_maps}
  \begin{split}
  f_\theta(x) &= U_\theta(x)\dyad{0^n}{0^n}U_\theta(x)^\dagger
  \\
  g_\eta(\tilde{x}) &=
  V_\eta(\tilde{x})\dyad{0^n}{0^n}V_\beta(\tilde{x})^\dagger
  \end{split}
\end{align}
where $U_\theta(x)$, $V_\eta(\tilde{x})$ are parameterized quantum circuits. Then
\begin{align}
  d(x, \tilde{x})^2 = 2 -
  \mathcal{S}_{\theta, \eta}(x, \tilde{x})
\end{align}
where the similarity measure
\begin{align}
  \label{e:full_measurement}
  \begin{split}
  \mathcal{S}_{\theta, \eta}(x, \tilde{x}) &= \Tr{f_\theta(x)^\dagger g_\eta(\tilde{x})} \\
  &=
  \abs{\expval{U_\theta(x)^\dagger V_\eta(\tilde{x})}{0^{\otimes
  n}}_H}^2
  \end{split}
\end{align}
is nothing but the overlap between states $U_\theta(x)\ket{0^n}$ and $V_\eta(\tilde{x}) \ket{0^n}$. There are multiple methods to efficiently compute \autoref{e:full_measurement} using a quantum device. It can be measured experimentally by first preparing the state $U_\theta(x)^\dagger V_\eta(\tilde{x}) \ket{0^n}$ and then computing the probability that all $n$ qubits measure to $0$ as shown in \autoref{fig:all circuits}(b). Alternatively, one can prepare the states $U_\theta(x) \ket{0^n}$, $V_\eta(\tilde{x}) \ket{0^n}$ in two separate $n$-qubit registers and performing a SWAP test with an ancilla qubit. The latter scheme trades circuit depth for width.

\subsection{Generalized similarity learning}\label{sec:gsimlearn}

In the previous scenario, we formulated the picture where data in each individual space is mapped by a unique embedding to a Hilbert space. The quantum feature maps are hitherto defined by applying unitaries to a standard $n$-qubit state $\dyad{0^n}{0^n}$. We will explore two variations of \autoref{e:full_measurement}, which by slightly tweaking, generalizes the previous setting.

Consider first a simple variant of the SWAP test method. Recall that in this setup, the feature maps $f_\theta(\cdot)$, $g_\eta(\cdot)$ act on separate registers of $n$ qubits each. For $m \le n$, perform the SWAP test only on the first $m$ qubits of each register to obtain

\begin{align}
  \mathcal{S}_{\theta, \eta}^m(x, \tilde{x}) := \Tr\bigl[f_{\theta, m}(x)^\dagger g_{\eta, m}(\tilde{x})\bigr], \label{eq:swap partial}
\end{align}
where the (mixed) states
\begin{align}
  f_{\theta, m}(x) = \Tr_{n-m}\bigl[{f_\theta(x)}\bigr]\\
  \ g_{\eta, m}(\tilde{x}) = \Tr_{n-m}\bigl[g_\eta(\tilde{x})\bigr]
\end{align}
are the partial trace of the original density matrices over the last $n-m$ qubits, such a circuit is shown in \autoref{fig:all circuits}(d). When $m<n$, $f_{\theta, m}$ encodes classical data into an $m$-qubit system by applying a possibly non-unitary CPTP map to the initial state $\dyad{0^m}{0^m}$.

The second tweak comes by slightly changing modifying how
\autoref{e:full_measurement} is measured without SWAP test. An alternative for
calculating expectation value, as mentioned previously, is to apply
$U_\theta(x)^\dagger V_\eta(\tilde{x})$ to $\ket{0^n}$ and measure
the probability that all $n$ qubits return $0$. Instead, we will now only
inspect the first $m$ qubits to obtain another modified similarity
\begin{align}
  \tilde{\mathcal{S}}_{\theta, \eta}^m(x, \tilde{x}) &= \sum_{i \in \{0, 1\}^{n-m}}\abs{\braket{0^m i}{\phi_{x, \tilde{x}}^{\theta,\eta}} }^2\\
  &=\Trace{\left\{\dyad{ 0^m}{0^m} \left(\Tr_{n-m} \left[\ket{\phi^{\theta,\eta}_{x, \tilde{x}}}\bra{\phi^{\theta,\eta}_{x, \tilde{x}}}\right]\right)\right\}}
\end{align}
where $\ket{\phi^{\theta,\eta}_{x, \tilde{x}}}=U_\theta(x)^\dagger
V_\eta(\tilde{x}) \ket{0^n}$. This circuit is shown in \autoref{fig:all
circuits}(c).

\begin{figure}[h!]
  \includegraphics[width=\linewidth]{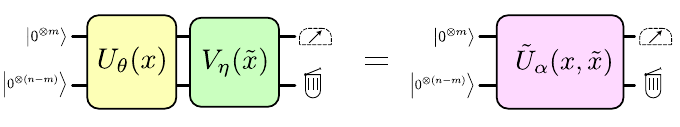}
  \caption{Mapping $x$ and $\tilde{x}$ using $U_\theta$ and $V_\eta$ can be generalized using a generic pair embedding $\tilde{U}_\alpha$ }\label{fig:effective}
\end{figure}

One way to think of this quantity is to regard the correspondence \[(x,
\tilde{x}) \mapsto U_\theta(x)^\dagger V_\eta(\tilde{x}) \ket{0^n}\] as an
embedding of \emph{pairs} $(x, \tilde{x})$. One might more suggestively
write$U_\theta(x)^\dagger V_\eta(\tilde{x})
=\tilde{U}_\alpha(x,\tilde{x})$ for some effective $\tilde{U}$ as shown
in \autoref{fig:effective}. Ideally, all similar pairs $(x,
\tilde{x}) \in \mathcal{E}_l$ should align perfectly with $\ket{0^n}$ and
dissimilar pairs $(x, \tilde{x}) \in \mathcal{D}_k$ should be orthogonal to
$\ket{0^n}$. Here, we do not lose generality by mapping it to $\ket{0^n}$ as
any other arbitrary state can be mapped back to $\ket{0^n}$ by a unitary which
then can be absorbed into the embedding parameterized unitary. Measuring only
some of the qubits effectively takes a partial trace over the remaining ones.
Note that unlike the previous formulation, $\tilde{\mathcal{S}}^m_{\theta,
\eta}$ is  need not be symmetric in $x$ and $\tilde{x}$, even when
$X=\tilde{X}$.

\subsection{Goodness of similarity learning}\label{subsec:goodness}

It is crucial to understand what it means for a pairwise functional maps $\mathcal{S}$ to be a ``good similarity function'' for a given learning problem.

\citet{balcan2008improved} proposed a new learning theory for similarity functions. This framework aims to generalize the learning techniques of kernels by relaxing some of the constraints we are currently interested in.  They give intuitive and sufficient conditions for a similarity function to guarantee performance. Essentially, a similarity function ($\mathcal{S}$) is $(\epsilon,\gamma,\tau)$-good if a $1-\epsilon$ proportion of examples are on average $2\gamma$ more similar to examples of the same class than to examples of the opposite class for a $\tau$ proportion of examples. This is explicitly proved for a general $\mathcal{S}$ which need not be a metric that satisfies positive semi-definiteness or symmetry. Given a $(\epsilon,\gamma,\tau)$-good $\mathcal{S}$, it can be shown that $\mathcal{S}$ can be used to build a linear separator in an explicit projection space that has a margin $\gamma$ and error arbitrarily close to $\epsilon$.

\begin{definition}[\citet{balcan2008improved}]\label{def1}
A  similarity function ($\mathcal{S}$) is $(\epsilon,\gamma,\tau)$-good function for a learning problem $P$ if there exists a (random) indicator function $\mathbf{R}(\mathbf{x})$ defining a (probabilistic) set of points such that the following conditions hold:
\begin{enumerate}
\item A $1-\epsilon$ probability mass of examples $(x,y(\mathbf{x}))$ satisfy
\begin{align*}\mathbb{E}_{\left(\mathbf{x}^{\prime}, y(\mathbf{x}^{\prime})\right) \sim P}\left[y(\mathbf{x}) y(\mathbf{x}^{\prime}) \mathcal{S}\left(\mathbf{x}, \mathbf{x}^{\prime}\right)\right ] \geq \gamma\end{align*}
\item $\operatorname{Pr}_{\mathbf{x}^{\prime}}\left(R\left(\mathbf{x}^{\prime}\right)\right) \geq \tau$.
\end{enumerate}
\end{definition}

Formally defined in \autoref{def1}, it can also be shown that the set of similarity maps defined by this is strictly larger than the set of metric maps and also includes non-positive semi-definite kernels and asymmetric maps. With the above definition, Ref\cite{balcan2008improved} proved the following theorem

\begin{theorem}[\citet{balcan2008improved}]\label{theorem1}
Let $\mathcal{S}$ be a $(\epsilon,\gamma,\tau)$-good similarity function for
a learning problem for a learning problem $P$. Let $\mathcal{L}=\left\{\mathbf{x}_{1}^{\prime}, \ldots, \mathbf{x}_{d}^{\prime}\right\}$ be a (potentially unlabeled) sample of $d=\frac{2}{\tau}\left(\log (2 / \delta)+8 \frac{\log (2 / \delta)}{\gamma^{2}}\right)$ landmarks drawn from $P$. Consider the mapping $\phi^{\mathcal{L}}: \mathcal{X} \rightarrow \mathbb{R}^{d}$ defined as following
\begin{align*}
\phi^{\mathcal{L}}(\mathbf{x})=\mathcal{S}\left(\mathbf{x}, \mathbf{x}_{i}^{\prime}\right), i \in\{1, \ldots, d\}.
\end{align*}

Then, with probability $1-\delta$ over the random sample $\mathcal{L}$ the
induced distribution $\phi^{\mathcal{L}}(P)$ in $\mathbb{R}^{d}$ has a
separator error at most $\epsilon+\delta$ relative to $L_1$ margin at least
$\gamma/2$.
\end{theorem}

\autoref{theorem1} states that, if enough data is available for the problem $P$, then for a $(\epsilon,\gamma,\tau)$-good similarity function, with high probability there exists a low-error (arbitrarily close to $\epsilon$) linear separator in mapped $\phi$-space. This framework of $(\epsilon,\gamma,\tau)$-good functions opens the door for us to evaluate the performance that one can expect a global linear separator to have, depending on how well a similarity function satisfies \autoref{def1}. We will see in \autoref{subsec:ND} that this method, at least numerically, can separate the classes as needed by definition.

\section{Discussion}\label{sec:discussion}

We will start with a qualitative discussion on similarity learning methods
introduced in \autoref{sec:multimetric}, following which we solve a toy
embedding explicitly to understand the intricacies of these methods. Next, we
explore a more complex example of learning the similarity between a
synthetic image set and abstract 2D space, and numerically show
that this learning method can learn salient encoded features of the images.

\subsection{Analytical Discussion}\label{subsec:adiscussion}

\begin{figure}[h!]
  \includegraphics[width=\linewidth]{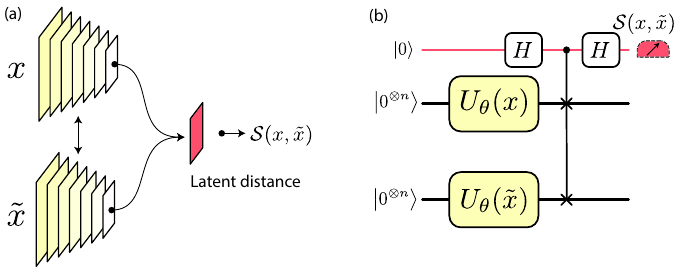}
  \caption{Classical Siamese network and its quantum counterpart. \textit{(a)} Classical Siamese network made up of two top and bottom networks that have same weights which project the input vectors to latent space after which we calculate distance on this space. \textit{(b)} Quantum feature maps defined in \autoref{e:feature_maps} with $g_\eta=f_\theta$ similar and their inner product being computed with ancilla.}\label{fig:siamese}
\end{figure}

By taking $X=\tilde{X}$ in metric learning setting, we reduce to the framework
of classical metric learning where the sought-after $d$ is assumed to be a
distance (pseudo-)metric in the usual sense. Thus, $d$ should satisfy the
following properties $d(x_1, x_2) \ge 0$ with equality when $x_1=x_2$; $d(x_1,
x_2) = d(x_2,x_1)$, and $d(x_1, x_3) \le d(x_1, x_2) + d(x_2, x_3)$ for all
triples $x_1,x_2,x_3 \in X$. One way to enforce these constraints is to seek a
distance function of the form\[d_\theta(x_1, x_2) = \| f_\theta(x_1) -
f_\theta(x_2)\|_Z\]where $f_\theta$ is some neural network mapping $X$ into
some latent Euclidean space $Z$. This construction, depicted in
\autoref{fig:siamese}, is called a \emph{Siamese neural
network}~\cite{bromley1993signature}. By taking $f_\theta(x) =
\dyad{\phi_\theta(x)}{\phi_\theta(x)}$ as a quantum feature map as in the
previous subsection, we have
\begin{align}
  \label{e:pythagorean}
  d_\theta(x_1, x_2)^2 = 2 - 2 K_\theta(x_1, x_2)
\end{align}
where $K_\theta(x_1, x_2) = \abs{\braket{\phi_\theta(x_1)}{\phi_\theta(x_2)}}^2$ is a parameterized quantum kernel\cite{Lloyd2020,Hubregtsen2021}. As noted in Ref.\cite[Appendix A]{Hubregtsen2021}, the training criterion \autoref{e:learning_criterion} implicitly guides the latent space embeddings $f_\theta$ to separate dissimilar pairs while compressing similar pairs together.

Next, in case of the similarity $\mathcal{S}^m$, as hinted in
\autoref{sec:gsimlearn}, when $m<n$, $f_{\theta, m}$ is a non-unitary CPTP map.
This thus enriches the family of feature maps at our disposal compared to the
previous case. Intuitively this can be understood as following, we prepare two
quantum states that map classical data $x$ and $\tilde{x}$ to their respective
states $\ket{x},\ket{\tilde{x}}$. In the previous multi-space metric learning
case, we required that (dis)similar elements be mapped (far)close to
each other in the Hilbert space $\mathcal{H}^n$ with dimension $n$. In
contrast, similarity defined in \autoref{eq:swap partial} requires that
dis)similar gets mapped (far)close to each other in the Hilbert space
$\mathcal{H}^{m} \subset \mathcal{H}^{n}$. This is similar to the
\textit{projective} SWAP test used in Ref\cite{lloyd2013quantum}.

We will now offer some heuristics for the more general similarity measure $\mathcal{\tilde{S}}^m$. It is easier to cluster points together in lower-dimensional spaces, which may help when similar raw data are far apart in their "native" spaces $X, \tilde{X}$. The trade-off is that separating points becomes hard. The number of qubits measured changes the relative difficulties of satisfying the training constraints.

Precisely, suppose one measures $m$ of the $n$ qubits. Then if $(x, \tilde{x})
\in \mathcal{E}_l$, the possible embeddings $\ket{\phi^{\theta,\eta}_{x,
\tilde{x}}}$ that satisfy the constraint $\mathcal{S}^m(x, \tilde{x}) = 1$ live
in a subspace of dimension $d_{\mathcal{E}_l} = 2^{n-m}$. This space is spanned
by all basis states $\{\ket{0i}, \ i \in \{0, 1\}^{n-m}\}$. On the other hand,
if $(x, \tilde{x}) \in \mathcal{D}_k$, the states $\ket{\phi^{\theta,\eta}_{x,
\tilde{x}}}$ that satisfy $\mathcal{S}^m(x, \tilde{x}) = 0$ constitute a
subspace of dimension $d_{\mathcal{D}_k} = 2^{n} - 2^{n-m}$. The ratio
$\zeta=\frac{d_{\mathcal{E}_l}}{d_{\mathcal{D}_k}}=\frac{1}{2^{m}-1}$
describes the relative difficulties of the training constraints. In the
scenario
where $m=1$, we have $\zeta=1$, \textit{i.e} we
provide equal volumetric space in the model to place similar and dissimilar
data points. In contrast, when $m=n$, $\zeta=\frac{1}{2^{n}-1}$ where we
provide larger space for dissimilar points to live. This gives us the ability
to tune the model space we are working with based on the data provided, for
instance for an highly imbalanced data set (when $\abs{l-k}>> 0 $), one can
choose $m$ based on if $\frac{l}{k}\approx \zeta$.

\begin{figure}[t]
  \includegraphics[width=\linewidth]{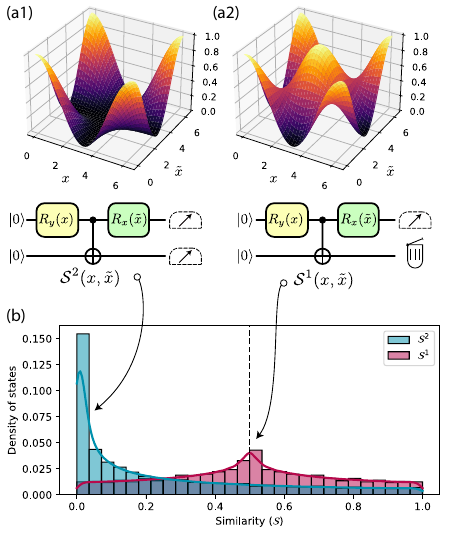}
  \caption{\textit{(a)} Quantum embedding and its corresponding similarity measure for $(a1)$ - $\mathcal{S}^2$ where we calculate the similarity of the joint quantum mapping of $(x,\tilde{x})\to \ket{00}$; $(a2)$ - $\mathcal{S}^1$ with $(x,\tilde{x})\to \sum_{i=\{0,1\}}c^{x,\tilde{x}}_i\ket{0\mkern 2mu i}$. In both $a1$ and $a2$, value of $1$ indicates $(x,\tilde{x})\in\mathcal{E}_l$ while $0$ indicates $(x,\tilde{x})\in\mathcal{D}_k$. \textit{(b)} Plot of the density of states of $\mathcal{S}^m$ for $m\in\{1,2\}$ as per \autoref{eq:dos}}\label{fig:toy-problem}
\end{figure}

We illustrate these considerations with a toy example. Consider the
following
simple two-qubit circuit -- $U(x,\tilde{x})=R_y(x,1)CNOT(0,1)R_x(\tilde{x},1)$
where $R_x(a, b)$ and $R_y(a,b)$ are the Pauli $X$ and $Y$ rotations by angle
$a$ on qubit $b$ -- and the
state $\ket{\phi^{\theta,\eta}_{x, \tilde{x}}}=\ket{\phi_{x,
\tilde{x}}}=U(x,\tilde{x})\ket{0^{\otimes 2}}$. $U$ can be interpreted as a
embedding of $x,\tilde{x}$ in a Hilbert space for a particular choice of
parameters. By traversing $0<x,\tilde{x}\leq 2\pi$, we can trivially look at
the entire embedding space accessible by the constant ansatz $U$. We can thus
look at the number of pairs of similar and dissimilar points that are
accommodated by this ansatz based on the measure $\mathcal{S}^m$ for various
$m$ (in this case $m\in\{1,2\}$), both of which is shown in
\autoref{fig:toy-problem}(a) as circuits. As shown in \cref{a:toy-problem} one
gets
\begin{align}
\mathcal{S}^2(x, \tilde{x})&=\cos^{2}{\left(\frac{x}{2} \right)} \cos^{2}{\left(\frac{\tilde{x}}{2} \right)}\label{eq:s2},\\
\mathcal{S}^1(x, \tilde{x})&=\frac{\cos{\left(x - \tilde{x} \right)}+\cos{\left(x + \tilde{x} \right)}}{4} + \frac{1}{2}\label{eq:s1}.
\end{align}

In \autoref{fig:toy-problem}(a), we plot $\mathcal{S}^1$ and $\mathcal{S}^2$
for all values of $x,\tilde{x}$. Intuitively, $\mathcal{S}$'s closeness to a
value of ($0$)$1$ indicates the paired points $x,\tilde{x}$ are (dis)similar to
each other. We see that in the case of $\mathcal{S}^2$ we have exactly 4 points
in the space that are maximally similar to each other. In contrast, a huge set
of points with an almost flat region have a similarity measure of $0$.
Comparatively, in $\mathcal{S}^1$, we have lifted a majority of this flat
region to support more points that are similar to each other. To better
understand this, we also look at the Density of States (DoS) of $\mathcal{S}$
for both the cases. The following defines this,
\begin{align}
D(\mathcal{S}):=\int \frac{\mathrm{d} x \mathrm{d} \tilde{x}}{(2 \pi)^{2}} \cdot \delta(\mathcal{S}-\mathcal{S}(x,\tilde{x}))\label{eq:dos}.
\end{align}
$D(\mathcal{S})$ essentially counts the number of points inside each fundamental unit of similarity. \autoref{fig:toy-problem}(b) shows the DOS, where we see a peak close to $0$, for $\mathcal{S}^2$ indicating that this measure allows for a high imbalance between similar and dissimilar points. In the case of $\mathcal{S}^1$, as discussed previously, we have a perfectly even split of volume between similar and dissimilar points with mid (dis)similarity measure being $0.5$.

\begin{figure}[t]
  \includegraphics[width=\linewidth]{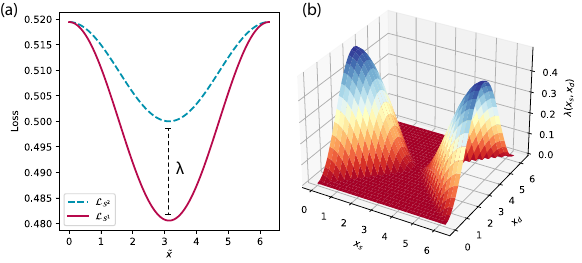}
  \caption{\textit{(a)}  Quality of similarity mapping measured using loss given in \autoref{eq:loss_toy} as a function of $\tilde{x}$. \textit{(b)} Improvement in similarity model $\lambda(x_s,x_d)$, from  \autoref{eq:loss_improv}}\label{fig:toy-problem-improv}
\end{figure}

Let us now place the above problem in the setting of \textit{retrieval} problem to gauge the ability of the respective family of similarity measures. Given two subspaces $X,\tilde{X}\in (0,2\pi]$, and $x_s,x_d\in X$, we strive to find some point $\tilde{X}$ that is most similar to $x_s$ but most dissimilar to $x_d$. This problem can be formulated as minimizing
\begin{align}
  \min_{\tilde{x}} \mathcal{L}_{\mathcal{S}}(\tilde{x},x_s,x_d)
\end{align}
where
\begin{align}
  \mathcal{L}_{\mathcal{S}}(\tilde{x},x_s,x_d)=\frac{1}{2}\left[(1-\mathcal{S}(x_s,\tilde{x}))^2+\mathcal{S}(x_d,\tilde{x})^2\right]^{\frac{1}{2}}\label{eq:loss_toy},
\end{align}

for some similarity measure $\mathcal{S}$ and loss function $\mathcal{L}$.
$\mathcal{L}(\tilde{x},\cdot,\cdot)$ measures the loss of how well the
similarity measure has performed. As a first illustration, we pick $x_s=0.3$
and $x_d=0.5$ and then calculate $\mathcal{L}(\tilde{x},x_s=0.3,x_d=0.5)$ using
the same embedding $U$. This is shown in \autoref{fig:toy-problem-improv}(a),
where the dotted lines use the measure $\mathcal{S}^{m=2}$ while the red line
uses $\mathcal{S}^{m=1}$. Since the quantum embedding map is the same for both
the cases, they both have the same optimal $\tilde{x}\approx\frac{\pi}{2}$. We
see that the loss function, which can be used as a surrogate to quantify how
good the similarity measure is in separating the optimal $\tilde{x}$, is much
lower/deeper in the case of $m=1$. Thus partial measurement of $m=1$ is able to
attain a better separation than $m=2$. To complete the analysis, we calculate
the quantity
\begin{align}
  \lambda(x_s,x_d)=\mathcal{L}_{\mathcal{S}^1}(\tilde{x}_{x_s,x_d}^{*},x_s,x_d)-\mathcal{L}_{\mathcal{S}^2}(\tilde{x}_{x_s,x_d}^{*},x_s,x_d)\label{eq:loss_improv},
\end{align}

where $\tilde{x}_{x_s,x_d}^{*}$ is the optimal $\tilde{x}$ for the pair $(x_s,x_d)$. $\lambda(x_s,x_d)$ quantifies the \textit{(dis)advantage} one gets by using $\mathcal{S}^1$ instead of $\mathcal{S}^2$. \autoref{fig:toy-problem-improv}(b) plots $\lambda$, where one sees that at worst case, we do not get any larger separation/better performance using partial measurement. But there are cases (as seen in \autoref{fig:toy-problem-improv}(a) and indicated by non zero value in (b)), where the (dis)similarity of $\tilde{x}$ is better in partial measurement.

\subsection{Numerical Discussion}\label{subsec:ND}
\begin{figure*}[!ht]
  \includegraphics[width=\linewidth]{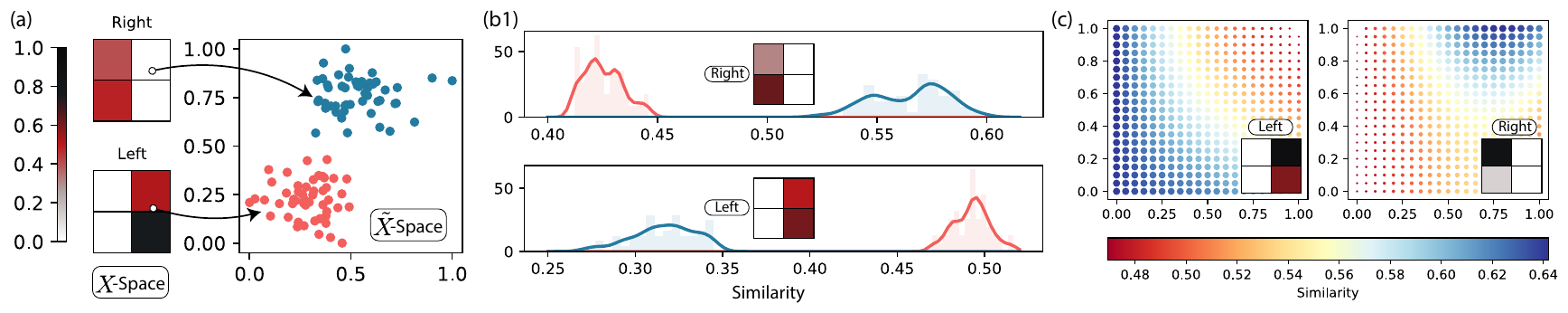}
  \caption{\textit{(a)} Left part shows the $X$-space data set, which consists of two sets of $2\times2$ images with one set having right pixels turned off and other set have left pixels turned off. The right part of the image shows $\tilde{X}$-space with red points being mapped to ``left'' images and blue points being mapped to ``right'' images in $X$-space. \textit{(b)} Density histogram of similarity of the out-of-sample inset $(b1)$ right image and $(b2)$ left image being compared to all in-sample red and blue points of $\tilde{X}$-space. \textit{(c)} Out-of-sample left and right image's similarity for a mesh of points in $\tilde{X}$-space with blue being the most similar and red being the most dissimilar. Size and color of the points are proportional to the similarity of the image in the inset.   }\label{fig:image-1}
\end{figure*}

\subsubsection{Experimental setup} \label{subsubsec:exp setup}
Unless otherwise stated, the following setup is adopted throughout the
numerical
experiments. We use the QAOA embedding\cite{lloyd2020quantum}, which typically
first encodes each input feature in the angle of a separate $R_X$ gate, then
creates entanglement using a combination of parameterized one-qubit $R_Y$  and
two-qubit $RZZ$ rotation. We use layers of size $2$ and qubits count of $4$.
We use the same basic ansatz to embed both features but allow
the underlying parameters to vary independently for each feature; in other
words, in~\eqref{e:feature_maps} we take $V_\eta = U_\eta$. Our training data
comes in the
form of subsets $\mathcal{E}_l, \mathcal{D}_k \subset X \times \tilde{X}$ of
similar and dissimilar pairs, respectively. By construction, our similarity
function $\mathcal{S}_{\theta, \eta}(x,\tilde{x})$ takes values in the interval
$[0, 1]$. We wish to find parameters $\theta, \eta$ such that
$\mathcal{S}_{\theta, \eta}$ ideally satisfies $\mathcal{S}_{\theta,\eta} = 1$
on $\mathcal{E}_l$ and $\mathcal{S}_{\theta, \eta} = 0$ on $\mathcal{D}_k$.  We
seek to minimize the cost function
\begin{align}
  \label{e:loss}
  L(\theta, \eta) = \sum_{(x,\tilde{x})} ( \mathcal{S}_{\theta, \eta}(x,\tilde{x})  - y_{x,\tilde{x}})^2,
\end{align}
where
\begin{align*}
  y_{x,\tilde{x}} = \begin{cases}
    1, & (x,\tilde{x}) \in \mathcal{E}_l\\
    0, & (x,\tilde{x}) \in \mathcal{D}_k.
  \end{cases}
\end{align*}

%Though we ideally seek the above condition, this leads to learning a measure,
%at the very least, maximizes
%\begin{align}
%\abs{\sum_{(x,\tilde{x})\in\mathcal{E}_l}
%\mathcal{S}_{\theta,\eta}(x,\tilde{x})-  \sum_{(x,\tilde{x})\in\mathcal{D}_k}
%\mathcal{S}_{\theta,\eta}(x,\tilde{x})},
%\end{align}
%\textit{i.e} the distance between the distributions formed by similarities of
%similar and dissimilar pairs are maximized (which would have ideally been 1
%and
%0 respectively). \textcolor{red}{This last claim doesn't obviously follow
%from~\eqref{e:loss}. Rather than trying to prove it I would omit it for now.}

Whereas most machine learning literature searches for the optimal parameters using some form of stochastic gradient descent, we employ the COBYLA optimizer \cite{powell2007view}.  This is a gradient-free method designed for noisy cost landscapes. Since it is expensive to evaluate the loss \autoref{e:loss} over the entire dataset, whenever it is required to evaluate the loss, we approximate it by a sum over a randomly chosen subset of the terms, normalized by the batch size of $80$, justification for this choice is given in \autoref{a:opt}.

\subsubsection{Experiment/Discussion}
\paragraph{Image data}: To better understand the capabilities of these learning models, we here show a full working flow of the algorithm. We synthetically generate $N=100$ $2\times2$ images as shown in \autoref{fig:image-1}(a) where the pixel values of images are filled from a uniform random distribution between $[0,1]$. We split the image data into two sets where we reset the right half of the pixels to $0$ for one set and the left half to $0$ in the other set. We label these two subsets as ``right'' and ``left''. This now forms the $X-$space of our data. For $\tilde{X}$-space, we generate $N=100$ points clustered into 2 sections (shown as red and blue in (a)) in $\mathbb{R}^2$. For the training data, we consider ``left''(``right'') images in $X$ to be similar to red(blue) points in $\tilde{X}$. Using this, we then generate our training points made up of $(x,\tilde{x},y)$ for all points $x\in X$ and $\tilde{X}\in \tilde{X}$ with $y$ being ($0$)$1$ for (dis)similar $(x,\tilde{x})$ pairs. We then start the learning process to find the optimal parameters ($\theta^*,\eta*$) that minimize the loss given in \autoref{e:loss}. We use $\mathcal{S}^{m=2}$ as the similarity model.

\paragraph{Performance}: Once we have the optimal angles, there are multiple ways to test the performance of this model. First, we generate a new random ``left'' ($x^l$) and ``right'' ($x^r$) image and calculate the following four quantities - $\mathcal{S}(x^l,\tilde{x}_i^{red})$, $\mathcal{S}(x^r,\tilde{x}_i^{red})$, $\mathcal{S}(x^l,\tilde{x}_i^{blue})$, $\mathcal{S}(x^r,\tilde{x}_i^{blue})$ where $\tilde{x}_i^{red/blue}$ is the $i^{th}$ red/blue point in $\tilde{X}$-space that was used for training. We plot the histogram/density of these points in \autoref{fig:image-1}(b), where we plot the similarity value of test image shown in inset with red points (shown as red line) and blue points (shown as blue line) for (b1) $x^r$ and (b2) $x^l$. We see that in both cases, the network successfully separates the similarity values of how close the left/right image is to the blue/red points depicted by the peaks in each distribution. Second, we see that $x^r$ is closer (higher similarity value) to blue points than to red points while the case is reversed for the left value. This is consistent with how we expect the model to behave. This shows that the model has not only trivially \textit{learned} what it means to be left image and what it means to be the right image, but how this feature associates to a different subspace - $\tilde{X}$. This brings us to the discussion in \autoref{subsec:goodness}, where we introduced the concepts of $(\epsilon,\gamma,\tau)$-good similarity function. The figure shows that the average distance between samples in classes is well separated for their corresponding classes in $X$ space. Numerically we verified that this is not just true for a subset of points in $X$, but for all points in $X$.

Another gauge of the model performance is to look at where in the
$\tilde{X}$-space does our model place a given point in $X$-space. To this end,
we chose another random $x^{l/r}$ and calculated its similarity on a uniform
grid space in $\tilde{X}$. This is shown in \autoref{fig:image-1}(c), where the
color and size of each grid point corresponds to the learned similarity measure
of the image in the inset. We see that the most similar points in $\tilde{X}$
for the left image (indicated by blue color) are around the lower left, while
for the right, it is around the upper right corner. This is indicative of the
initial mapping we learned from where the points in $\tilde{X}$ space form
clusters in the left bottom, and top right parts of the space.

\begin{figure}[h]
  \includegraphics[width=\linewidth]{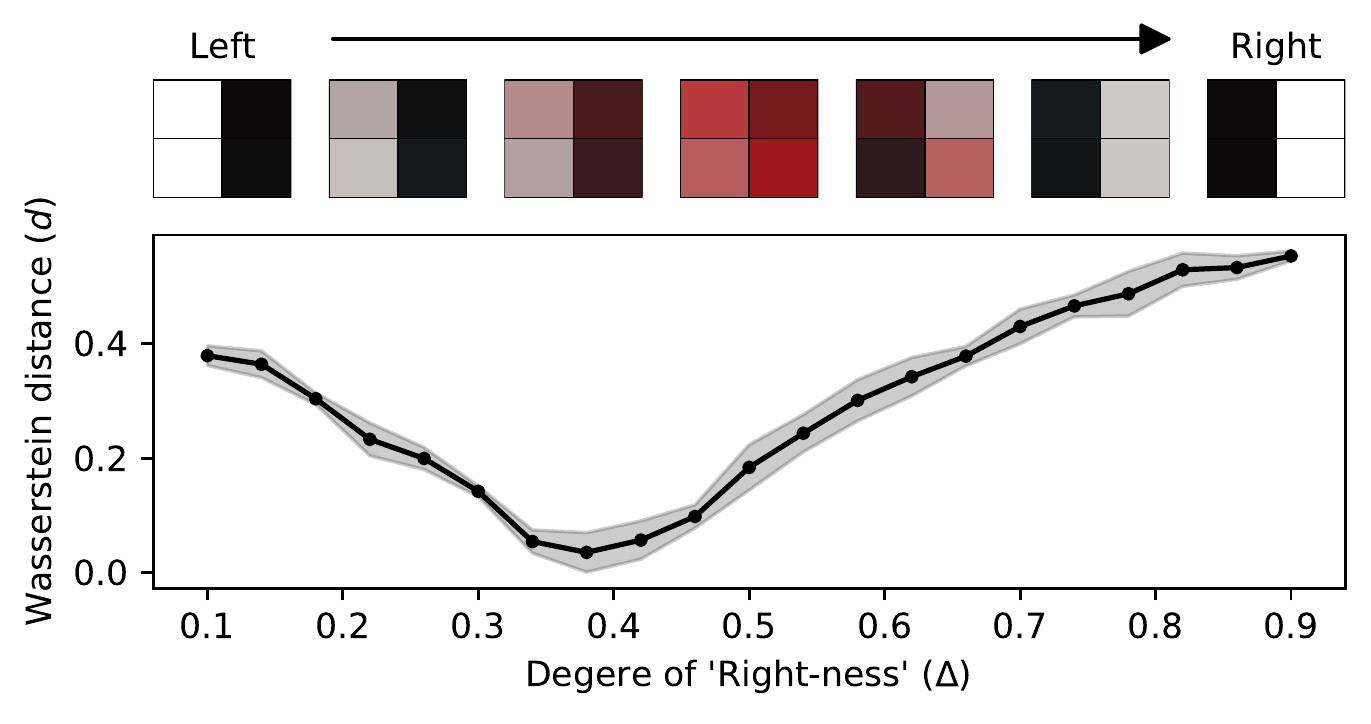}
  \caption{\textit{(top)} Sample images for values for equally spaced
    $\Delta$ between $[0,1]$. \textit{(bottom)} Average Wasserstein
    distance
    $(d)$ between the distribution of each image $x_\Delta$ and its
    variance as
    a function of $\Delta$. A low value of $d$ indicates that the image is
    equally (dis)similar to both blue and red points in $\tilde{X}$ while a
    high value indicates that it is closer to one class of points in
    $\tilde{X}$. This model is trained only at the end points in $\Delta$
    \textit{i.e} ``left'' and ``right'' images only.
  }\label{fig:image-transition}
\end{figure}

\paragraph{Generalizability of model}: To answer the question about the generalizability of the learned model, we now generate out-of-sample images given by the following $2\times2$ pixel value
\begin{align}
\begingroup % keep the change local
\setlength\arraycolsep{3pt}
\begin{matrix}
x_{\Delta}=
\begin{array}{c|c}
  \mathit{X}_{\Delta}& 1-\mathit{X}_{\Delta} \\
  \hline
  \mathit{X}_{\Delta}& 1-\mathit{X}_{\Delta},
 \end{array}
\end{matrix}
\endgroup
\end{align}
where $\mathit{X}_{\Delta}$ is a random variable generated from a truncated normal distribution  $0\leq \mathcal{\tilde{N}}(\Delta,\sigma=\epsilon)\leq1$ with mean $\Delta$ and variance $\sigma=\epsilon$, where $\epsilon$ is a relatively small number (in our case we choose $\epsilon=0.1$. Thus we have the two extremes $\Delta=0(1)$, where $x_{\Delta}$ belongs to left(right) image groups while it interpolates between left and right as a function of $\Delta$. This is shown in \autoref{fig:image-transition}(top).

We can quantify the network's performance in this scenario by looking at the
maximal separation between the similarity of $x_\Delta$ from the training
data's red and blue points. We do this by calculating the Wasserstein distance
($d$) between the distributions formed by these similarities. For example, in
\autoref{fig:image-1}(b1), we calculate the Wasserstein distance between the
red and blue distributions. We do this for $N=50$ different randomly generated
$x_\Delta$ images for each $0\leq\Delta\leq1$.
\autoref{fig:image-transition}(bottom) shows the result of such calculation
with error bars indicating the variance of the $50$ runs. Intuitively
$d(\Delta)$ measures the (in)distinguishably of $x_\Delta$ in $\tilde{X}$
space. Since in our mapping, left and right images are separable in $\tilde{X}$
space, we expect to see a high distance value at these extremes. As we approach
$\Delta=0.5$, the image becomes equally (dis)similar to both red and blue
points in $\tilde{X}$ and hence the distance between the respective similarity
distributions attain $0$. This is shown in
\autoref{fig:image-transition}(bottom). This is phonologically similar to the
ordered to disordered phase transition detected using classical neural network
in Ref\cite{shiina2020machine}. Thus by just training the model with the
extremities ($\Delta=0/1$), the network has learned to detect higher-order
features enabling us to detect unseen transitions. It shows that the final
achieved similarity scores are discriminative not just because of local
information about the images, rather a combination of both local and global
properties.

\section{Applications}\label{sec:app}

Although similarity measures can be used for a wide range of applications, we
choose three such applications to analyze and understand the technique. First,
we show the trivial example of how similarity learning can be used for
classification. Then, we illustrate a practical example of graph link
completion problem using similarity learning.
Finally, we show that ability to differentiate parameters in quantum circuits
efficiently gives us the ability to use the learned model as a generative model.

\subsection{Classification}
\begin{figure}[H]
  \centering
  \includegraphics[width=0.9\linewidth]{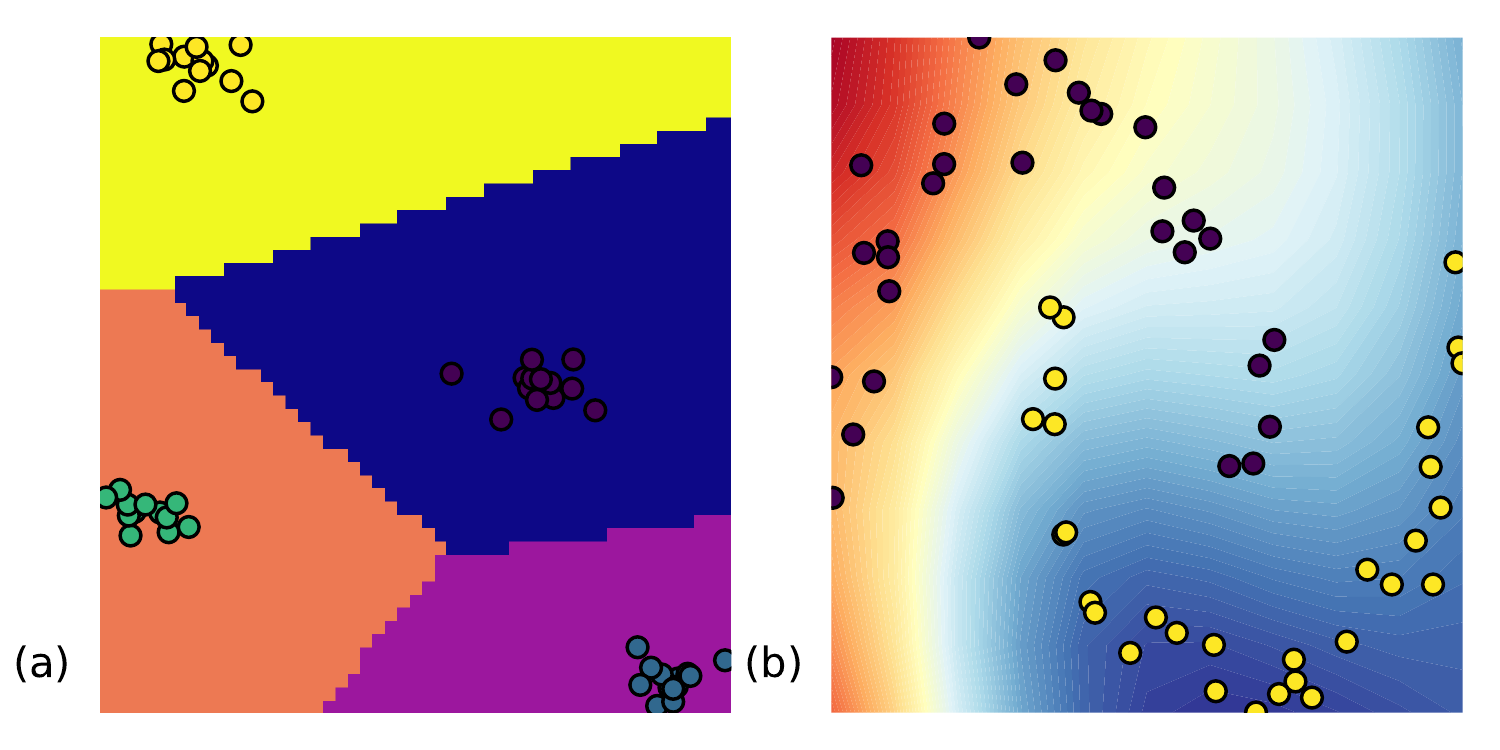}
  \caption{\textit{(a)} Multi-class classification by comparing the similarity with each class. Colors in points depict the training class while the background color depicts the predicted class.\textit{(b)} Moon-data set classification with the background color the probability of each point in mesh to belong to either class.}\label{fig:classification}
\end{figure}

Classification tasks and the notion of similarity are deeply connected. Suppose we are
given $m$ clusters of data \[C_1 = \{x_{1,i}\}_i, \ C_2 = \{x_{2, i}\}_i, \dots, C_m = \{x_{m, i}\}_i \subset \mathbb{R}^d.\] To classify a new data point $x \in \mathbb{R}^d$, we want to determine the class ``most similar" to $x$ - in other words, a \emph{fidelity classifier}\cite[Appendix C]{Lloyd2020}. Here, maximizing the separation of embedded data from different classes amounts to minimizing the empirical risk of the classifier for a linear loss function. An alternative method, which we do not pursue here, is to use the trained kernel in a support vector machine classifier as proposed by Hubregtsen et al~\cite{Hubregtsen2021}. Instead, it suffices to compare $x$ with a representative sample from each class. While the usual distance in $\mathbb{R}^d$ provides a naive notion of similarity, distance measures tailored to the data may perform better if the classes are irregularly shaped. In \autoref{fig:classification}, we show our \textit{GQSim} being used for $\mathcal{X}=\mathcal{\tilde{X}}=\mathbb{R}^2$ where (a) is multi-class classification and (b) is a more complex structure for similarity measure. In (a), we compare each point in the the space to single sample from each train class to do a one shot classification while in (b) we compute the similarity and normalize it to 1 for each point in space to get the probability of them being similar to yellow or purple family.

The image example discussed in
\autoref{subsec:ND} is a toy model of the following classical problem: given
$m$ clusters of images and crude information about the clusters, one can use
this information to encode the images in a different subspace $\tilde{X}$. This
information is termed \emph{side information}, and it appears naturally in many
applications. For example, when classifying images containing \{house, dog,
cat\}, we know that even though they belong in discrete categories, the
distance between categories dog-cat is closer than dog/cat-house. Another
example of side information is categorizing the rating of
a given movie in \{1,2,3\}. Even though each movie has a category, one has the
information that the categories are ordered \emph{i.e} $1<2<3$. This
information can be encoded in $\tilde{X}$ space by manually choosing clusters
for $\{1,2,3\}\to \{c_1,c_2,c_3\}$, say in $\mathbb{R}^1$, such that center of
$c_1<c_2<c_3$. Thus, now learning the similarity between the movie space and
$\tilde{X}$ space and using this similarity to categorize movies has more
implied structure built into it than directly classifying them.

\subsection{Graph Completion}
Link prediction\cite{bellet2013survey} is an important aspect of network analysis and an area of key research. Graph completion problem\cite{bai2019quantum} is a subset of link prediction, where it is assumed that only a small sample of a large network (e.g., a complete or partially observed sub-graph of a social graph) is observed, and one would like to infer the unobserved part of the network. To formalize the setting, we assume there is a true uni directed unweighted graph $\mathcal{G}=(\mathcal{V},\mathcal{E})$ on $n=|\mathcal{V}|$ distinguishable nodes with its own attributes $\mathbf{v}_i$ for $1\leq i \leq n$ with an adjacency matrix $\mathbf{A}\in \{0,1\}^{n\times n}$ where $0$ denotes an absent edge and $1$ denotes a connected edge. We then assume that only a partially observed adjacency sub-matrix $\mathbf{O}\in \{0,1\}^{m\times m}$ with $1\leq m \leq n$ induced by a sample sub-graph $\mathcal{G}^\prime =(\mathcal{V},\mathcal{E}^\prime)$ with $\mathcal{E}^\prime \subset \mathcal{E}$, of original graph is given. We are now interested in predicting the complete adjacency matrix $\mathbf{A}$ based on the partially observed sub-matrix $\mathbf{O}$ using the \textit{learned} similarity measures of the node attributes $\mathbf{v}_i$. For example, in a social community network, the nodes might represent the individuals, and the link of nodes represents the relations between the individuals. The attributes for each node could include the person’s metadata like interest, age, location, occupation, etc. The more similar attributes of the two nodes are, the higher probability they have a relationship and are linked in the network.
\begin{figure}[h!]
  \includegraphics[width=\linewidth]{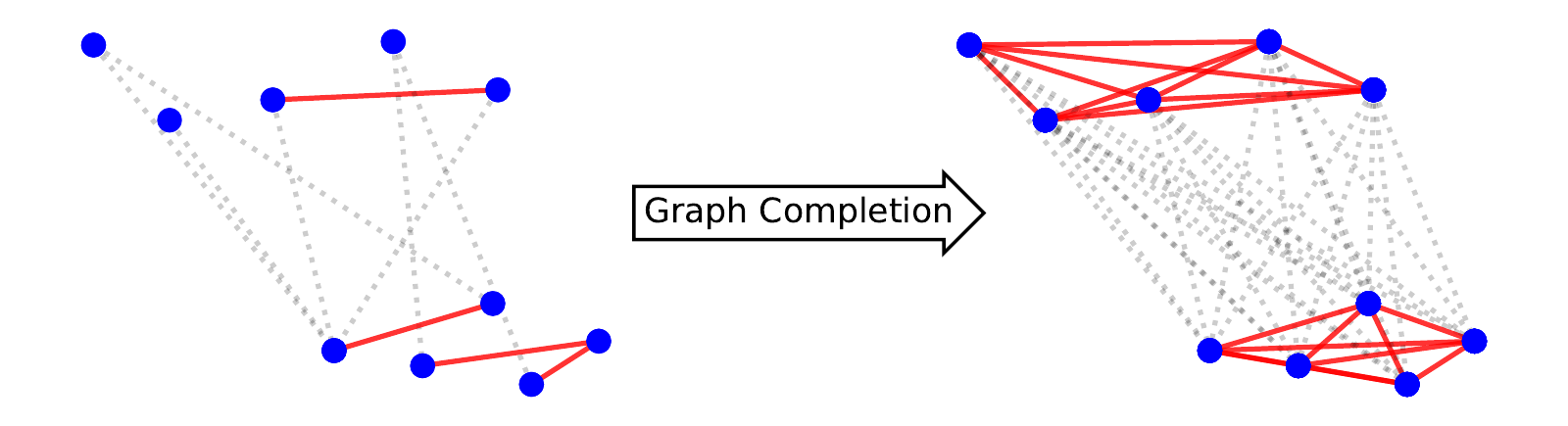}
  \caption{\textit{(left)} Input graph ($\textbf{O}$) with partial connection $\{0,1 \}\in\{\text{white dotted}, \text{red lines}\}$ are labelled \textit{(right)} Completed graph ($\textbf{A}$) where the edges are connected using the similarity algorithm.}\label{fig:graph-completion}
\end{figure}
As an example, in \autoref{fig:graph-completion} we will associate attributes
of each node with a $d=2$ dimensional data coming from $k=2$ unique
distributions $P_k$. Edges are then connected ($1$) to nodes having attributes
from the same distribution and not connected $(0)$ if their attributes
correspond to a different distribution. This defines the real hidden graph
$\mathbf{A}$. After this, we randomly select $\approx 10\%$ of the edges to
create the sub-graph $\mathbf{O}$. This is then fed into the algorithm to learn
the similarity between the nodes, which is then used to complete the graph.
Since we have chosen an uni-directed graph $\mathcal{G}$, connection between
$(v_i,v_j)\in \mathcal{V}$ is going to be the same between $(v_j,v_i)$. This
symmetry is forced in the model by setting $f_\theta=g_\eta$ of the learning
model, \textit{i.e.} we choose to embed both comparing points with the same
embedding. Clearly, for general directed multigraph, one needs to go beyond
symmetry and have the property that $(v_i,v_j)\neq (v_j,v_i)$. Moreover, in the
case of a directed multigraph, one needs to allow for nontrivial $(v_i,v_i)$ to
indicate loops in the graph potentially. For these cases, one may opt to use
the generalized similarity learner to embed the points in different embedding
and trace out parts of the Hilbert space.

The figure shows the graph nodes being plotted in a 2D space, with each node placed corresponding to its attribute. Red lines indicate the connected nodes, while dotted lines indicate no connection. Missing lines in the left figure correspond to the lack of information. The right figure shows the predicted complete matrix. As seen from the figure, the points from the same spatial cluster are connected while the nodes with attributes belonging to different $P_k$ are disconnected. In the experiments where we have $P_k$ as Gaussian with random spread $\in[0.5,1.5]$, a reduced graph with edges as little as $1\%$ reproduces the final graph with $100\%$ accuracy.

\subsection{Generative model}

\begin{figure}[h!]
  \includegraphics[width=0.6\linewidth]{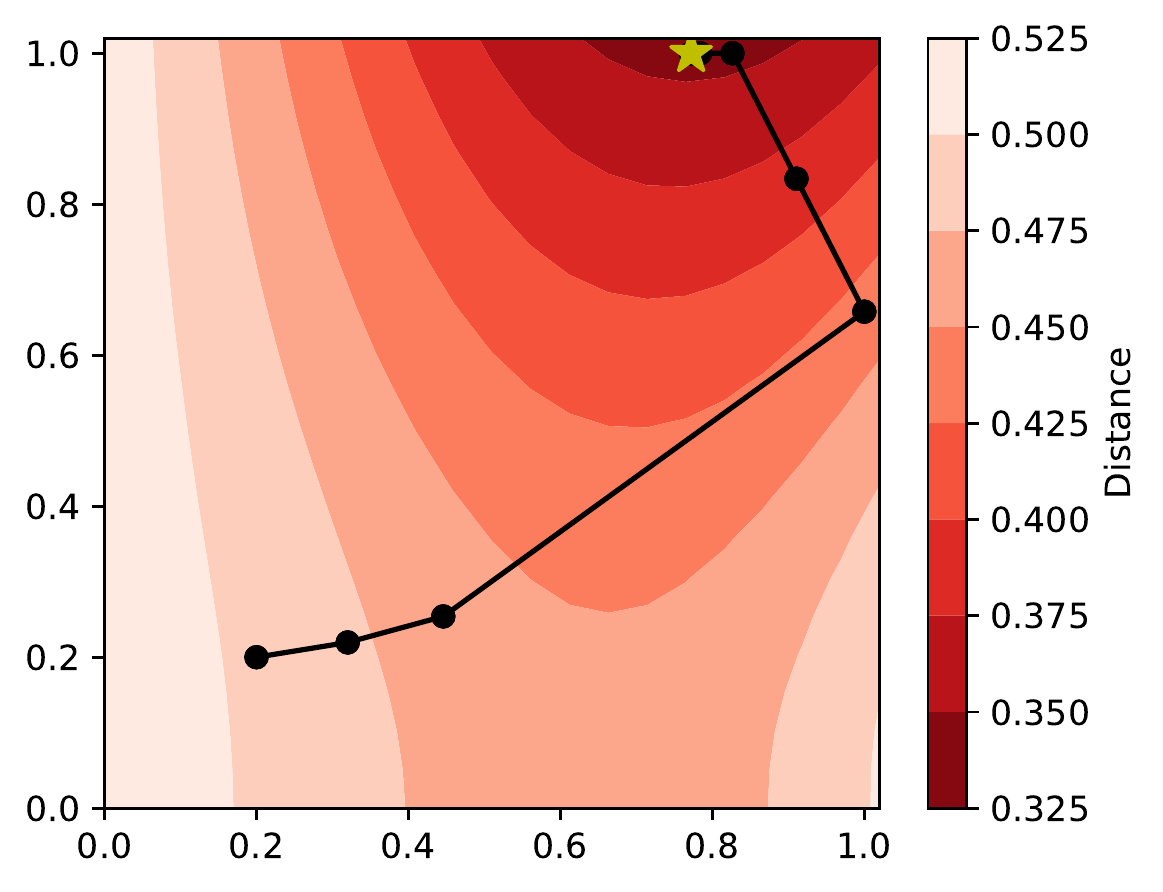}
  \caption{Cost landscape in $\tilde{X}$ space for the optimization problem in \autoref{eq:gen} with a particular ``right'' image in $X$. Path taken for optimization is shown in black line with the optimal data is marked with yellow star.  }\label{fig:generative}
\end{figure}

Given a learned similarity measure between two different spaces, one can now use support points to generate new data in the other space. More concretely, given a learned similarity measure $\mathcal{S}$, and a support point $x_s\in X$, one can generate an unseen data point $\tilde{x}_s$ in $\tilde{X}$ using the following optimization problem
\begin{align}
  \min_{\tilde{x}} 1- \mathcal{S}(x_s,\tilde{x}).\label{eq:gen}
\end{align}

Such tasks occur naturally in many scenarios; for example, in the case of language translation between two languages, one can generate new unseen similar sentences in another language based on learned similarity measure. Despite being an optimization problem in feature space, this is efficient in our quantum case. This is because features are directly encoded as parameters in our PQCs and thus can be efficiently differentiated\cite{benedetti2019parameterized}. To illustrate this, we again invoke the example discussed in \autoref{subsec:ND}. Given an image, we solve \autoref{eq:gen} to find the closest point in $\mathbb{R}^2$ for a given image of type ``right''. We show this in \autoref{fig:generative}, where we plot the cost function of this minimization problem and the minimization steps.

\section{Summary}\label{sec:summery}

We have considered a generalization of similarity learning techniques in the
quantum setting. PQCs express
similarity
functions with a richer set of properties like (a)symmetry, intransitivity, and
even multi-space metrics. Similarity learning boils down to
learning \emph{pair wise} similarity by embedding the input features in Hilbert
space. We
illustrate the effect of using partial measurements and their use in modeling
imbalanced data. Using a synthetic image data set with left and right pixels
blocked, we learn the similarity of these images $w.r.t$ arbitrary points in
$\mathbb{R}^2$ space. This example is also used to illustrate the generalizing
capability of these learned models. We use the model to detect phase transition
from left-like image to right-like image. Finally, we show three applied use
cases where learned similarity can be used. Classification, being a trivial use
case of learning similarity, can be augmented to encode given side-information
about the data using the multi-space property of generalized similarity
learning. Graph link completion problem can be rephrased as a similarity
learning problem, which we numerically showed for a simple example. Finally, we
demonstrate how similarity models can be used as generative models to generate
unseen data in the complementary space, given corresponding data in the
original space. Nevertheless, there are still critical challenges regarding the
choice of the PQC embedding one chooses for the given data.
\begin{figure}[t]
  \includegraphics[width=\linewidth]{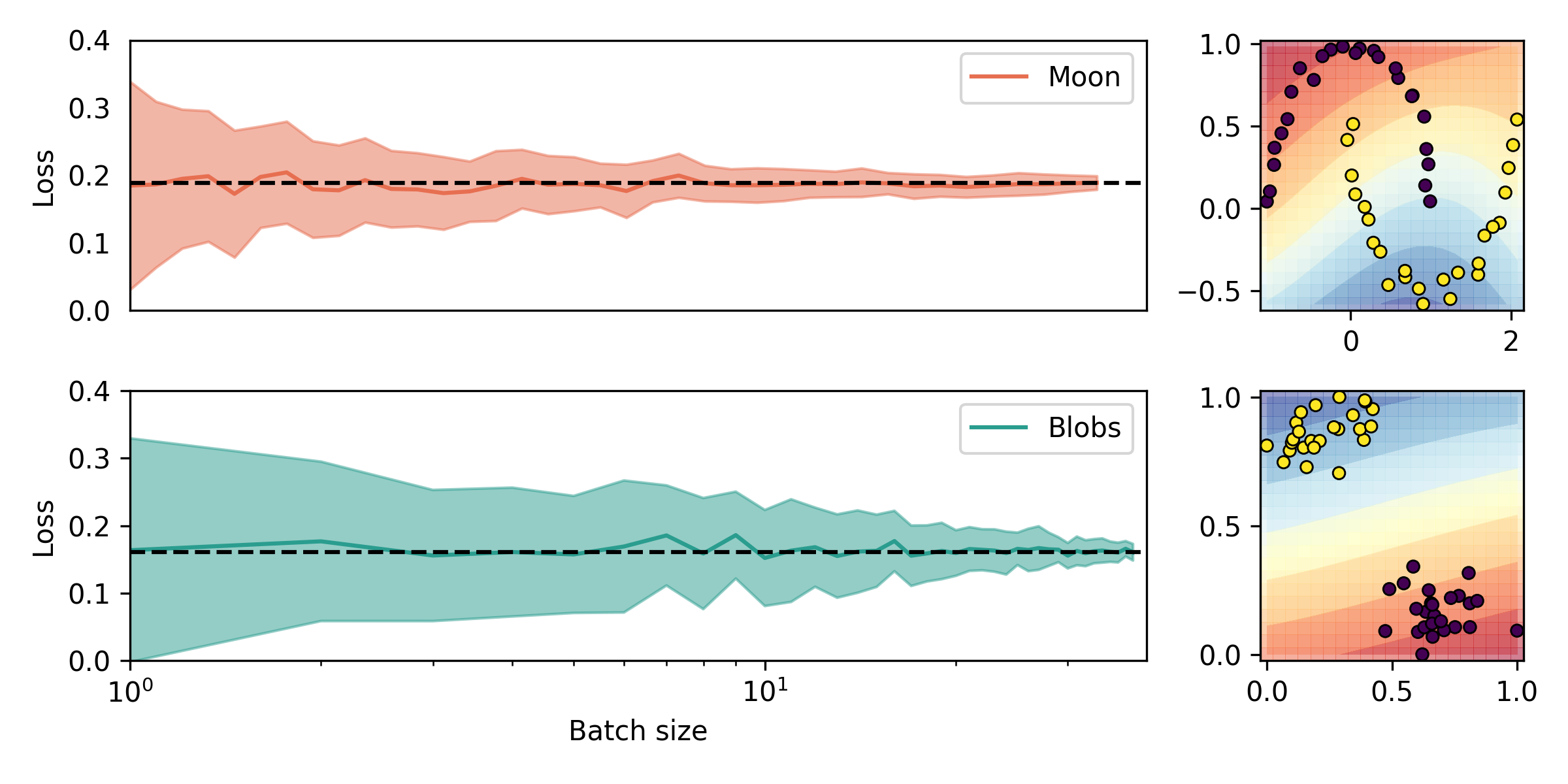}
  \caption{(Normalized) loss after training vs (log) batch size for $100$ runs of each batch size at the global optimum for \textit{(a)} blob data set and \textit{(b)} moon data set }
  \label{fig:loss_batch}
\end{figure}

\subsection{Acknowledgments}

The authors thank Jack Baker for insightful discussions.

\appendix
\section{Analytical details of toy-problem}\label{a:toy-problem}

We here outline the details of setup discussed in \autoref{subsec:adiscussion}.
As discussed, one can think of the PQC in \autoref{fig:toy-problem} (a1/a2) in
two equivalent ways - 1. $U(x)=R_y(x,0);V(\tilde{x})=CNOT(0,1)R_x(\tilde{x},1)$
(or move the $CNOT$ to $U$) or 2.
$\tilde{U}(x,\tilde{x})=R_y(x,0)CNOT(0,1)R_x(\tilde{x},1)$. First one
corresponds to individual mappings while the second one takes the pair values
and map it to a point in Hilbert space, both of which are equivalent views.
The matrix form of
$\tilde{U}$ is given by
\begin{widetext}
\begin{align}
\begingroup
  \renewcommand{\arraystretch}{1.5}
  \tilde{U}(x,\tilde{x})=\left[\begin{matrix}\cos{\left(\frac{x_{1}}{2} \right)} \cos{\left(\frac{x_{2}}{2} \right)} & i \sin{\left(\frac{x_{1}}{2} \right)} \sin{\left(\frac{x_{2}}{2} \right)} & - i \sin{\left(\frac{x_{2}}{2} \right)} \cos{\left(\frac{x_{1}}{2} \right)} & - \sin{\left(\frac{x_{1}}{2} \right)} \cos{\left(\frac{x_{2}}{2} \right)}\\i \sin{\left(\frac{x_{1}}{2} \right)} \sin{\left(\frac{x_{2}}{2} \right)} & \cos{\left(\frac{x_{1}}{2} \right)} \cos{\left(\frac{x_{2}}{2} \right)} & - \sin{\left(\frac{x_{1}}{2} \right)} \cos{\left(\frac{x_{2}}{2} \right)} & - i \sin{\left(\frac{x_{2}}{2} \right)} \cos{\left(\frac{x_{1}}{2} \right)}\\\sin{\left(\frac{x_{1}}{2} \right)} \cos{\left(\frac{x_{2}}{2} \right)} & - i \sin{\left(\frac{x_{2}}{2} \right)} \cos{\left(\frac{x_{1}}{2} \right)} & - i \sin{\left(\frac{x_{1}}{2} \right)} \sin{\left(\frac{x_{2}}{2} \right)} & \cos{\left(\frac{x_{1}}{2} \right)} \cos{\left(\frac{x_{2}}{2} \right)}\\- i \sin{\left(\frac{x_{2}}{2} \right)} \cos{\left(\frac{x_{1}}{2} \right)} & \sin{\left(\frac{x_{1}}{2} \right)} \cos{\left(\frac{x_{2}}{2} \right)} & \cos{\left(\frac{x_{1}}{2} \right)} \cos{\left(\frac{x_{2}}{2} \right)} & - i \sin{\left(\frac{x_{1}}{2} \right)} \sin{\left(\frac{x_{2}}{2} \right)}\end{matrix}\right],\label{eq:a:u}
\endgroup
\end{align}
\end{widetext}
from which we see that
\begin{align}
\abs{\expval{\tilde{U}(x,\tilde{x})}{00}}^2=\cos^{2}{\left(\frac{x_{1}}{2} \right)} \cos^{2}{\left(\frac{x_{2}}{2} \right)},
\end{align}
which is given in \autoref{eq:s2}. Similarly, we get \autoref{eq:s1} by computing
\begin{align}
\sum_{i\in\{0,1\}}\abs{\expval{\tilde{U}(x,\tilde{x})}{0i}}^2=\frac{\cos{\left(x - \tilde{x} \right)}+\cos{\left(x + \tilde{x} \right)}}{4} + \frac{1}{2},
\end{align}
which trivially from \autoref{eq:a:u}.
\begin{figure*}
  \includegraphics[width=\linewidth]{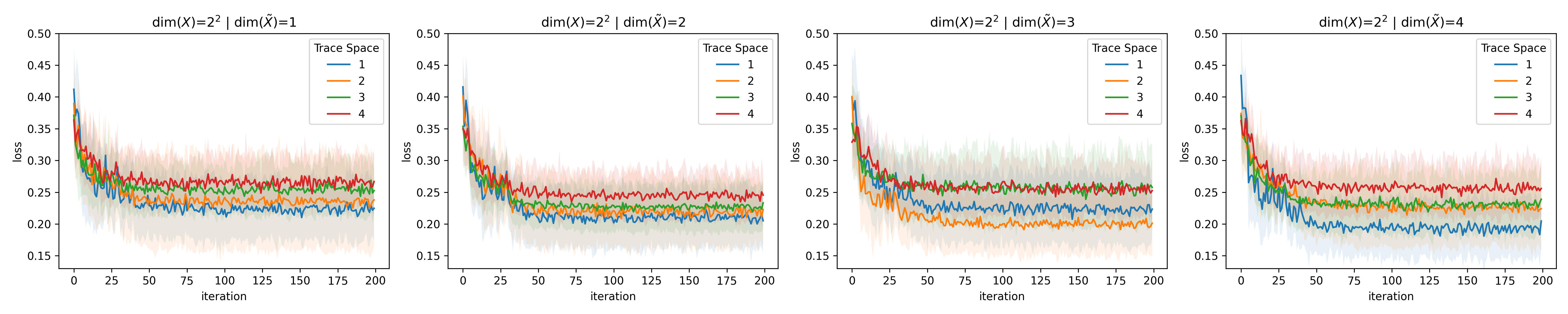}
  \caption{Loss vs iterations for different number of qubits measured indicated by Trace space where each plot referes to learning similarity between images of dimension dim$(X)$ and data points in dim$(\tilde{X})$}
  \label{fig:trace_loss_convergence}
\end{figure*}
\section{Optimization for learning}\label{a:opt}
Throughout the paper, as mentioned in \autoref{subsubsec:exp setup}, we have used the COBYLA\cite{powell2007view} optimization scheme for learning the similarity. Because of the costly function evaluations required to compute pair-wise terms in \autoref{e:loss}, we apply the heuristics of stochastic batching to reduce the number of evaluations. The heuristics are as follows:
\begin{itemize}
\item We pick $N^{b}_i/2$ random pairs of points from each similar/dissimilar pairs from the training set for $i^{th}$ iteration making the set $\mathcal{X}_N^i=X_{N^{b}_i}\times\tilde{X}_{N^{b}_i}$.
\item We compute $L_i(\theta, \eta) =\frac{1}{N^{b}_i} \sum_{(x,\tilde{x})\in \mathcal{X}_N^i} ( \mathcal{S}_{\theta, \eta}(x,\tilde{x})  - y_{x,\tilde{x}})^2$
\item We use this $L_i$ in the COBYLA optimization routine until convergence is achieved.
\end{itemize}
We caution that this ``stochastic'' form of COBYLA is similar to stochastic gradient descent techniques\cite{bottou2010large}, but has no theoretical backing yet in the literature. Having said that, this heuristic, although employed for reducing the computational time, could offer potential benefits that exist in stochastic gradient descent. Primarily, the stochastic batching heuristic could be used to reduce the number of function evaluations required for learning and push towards the global minimum much faster than non-stochastic, where we have lots of local maxima/minima. One can think of this as approximating the "effective" manifold of the cost function having an error bar. To this end, it is essential to understand what numerical choice of $N^{b}_i$ is best. Naively one could assume that $N^{b}_i$ is indeed data-dependent as the features in the data could make the landscape more complicated. To understand this better, we calculate the effect of $N^{b}_i$ on the cost function by computing the average of $L_i$ over a range of $N^{b}_i$ values for various runs. We then plot the average cost function over the range of $N^{b}_i$ for two different similarity learning problems - Classifying two clustered blob data set and two moon data set, which is shown in the \autoref{fig:loss_batch}.

We see that, at least in the case of blob vs. moon, where one would assume learning blob is more straightforward than learning similarity between moon, the batch size dependence on the variance of cost manifold is almost the same. We also observe that around batch size $=10^2$, the variance in cost manifold $\approx 10^{-2}$ for both blob and moon data set. This is because the cost function is very smooth, and the batch size is big enough to approximate the exact cost manifold, despite the actual batch size cost manifold having $\approx 1500$ pairs. We thus use a batch size of $\approx 81$ throughout the numerical experiments in the paper.

\section{The effect of partial measurements}
\begin{figure}[h]
\includegraphics[width=0.8\linewidth]{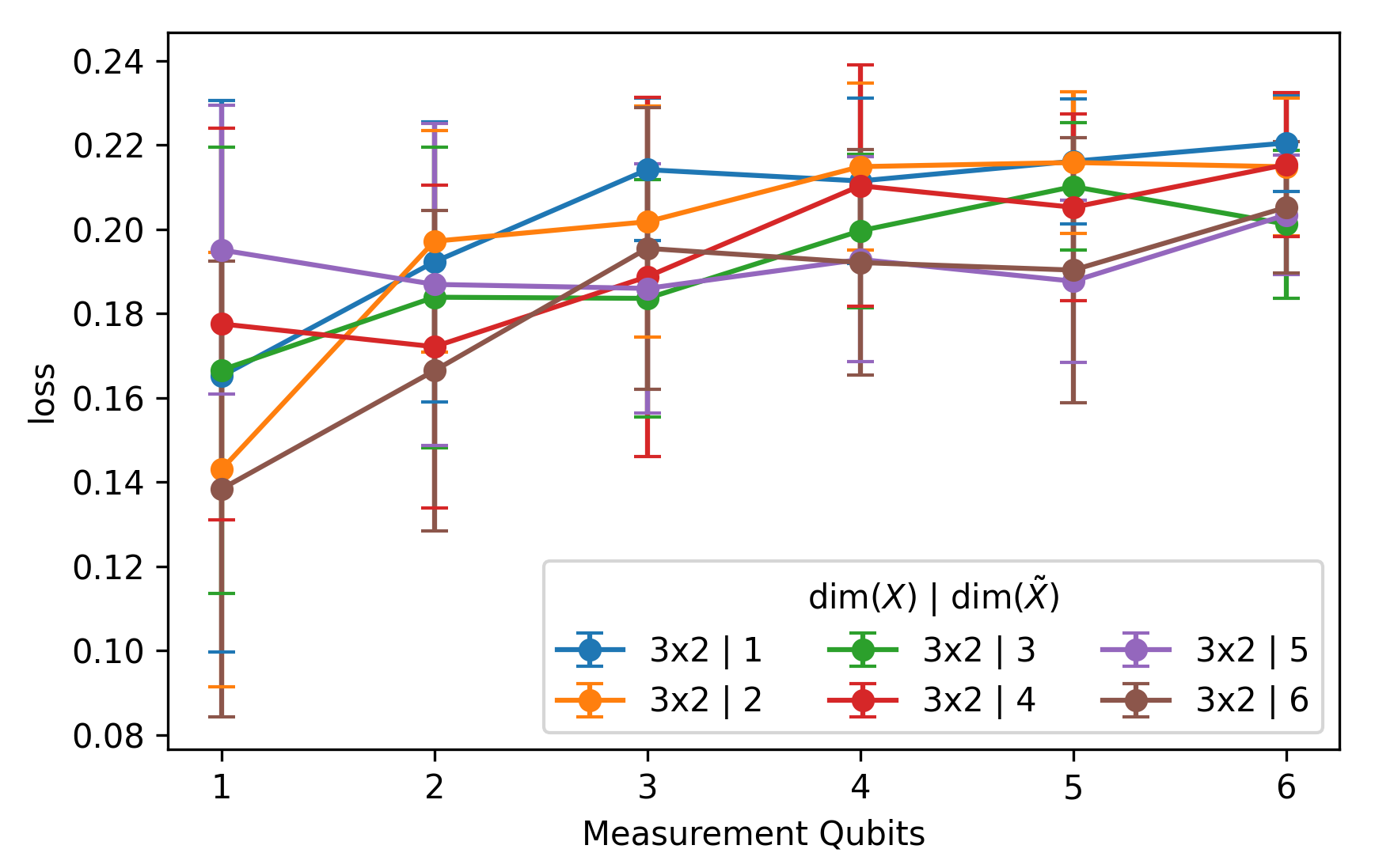}

\caption{Variance of optimal loss value for various problem instances shown in legend plotted along with the qubit space we are tracing}
\label{fig:trace_min_loss}
\end{figure}
We present some preliminary data on the effect of measuring only some qubits in the image data experiment of \autoref{subsubsec:exp setup}. For each $k = 1, 2, 3, 4$, generate two separated clusters, labeled ``red'' and ``blue'', of points in the $k$-dimensional unit cube $\tilde{X}= [0, 1]^k$; the preceding discussion considered the case $k=2$. As before we try to learn the associations (``Right'', \{blue cluster\}), (``Right'', not \{red cluster\}), (``Left'', \{red cluster\}), (``Left'', not \{blue cluster\}). We then run $100$ instances of randomly generated synthetic data of this form and run the learning algorithm for all the allowed trace qubit dimension. \autoref{fig:trace_loss_convergence} shows the loss as a function of training iterations for each dimension of $\tilde{X}$.

Next we run the training multiple times for each dimension of $\tilde{X}$ and for each choice of qubits to be measured. The average (mean) minimum training losses encountered during each training is shown in \autoref{fig:trace_min_loss}, together with their variance. Overall there appears to be a slight degradation in training quality as more qubits are measured. Note that for a given problem (denoted by color in plot ), all the trace experiments are done with the same data-set. We see that smaller Hilbert space seem to attain lower average cost value compared to measuring all qubits. We see that the variance in the cost drastically decreases as well. Both these phenomenon point us in future directions to explore for further understanding the effects of partial measurements in learning models.
% \clearpag\
\bibliography{paper}

%apsrev4-2.bst 2019-01-14 (MD) hand-edited version of apsrev4-1.bst
%Control: key (0)
%Control: author (8) initials jnrlst
%Control: editor formatted (1) identically to author
%Control: production of article title (0) allowed
%Control: page (0) single
%Control: year (1) truncated
%Control: production of eprint (0) enabled
\begin{thebibliography}{31}%
\makeatletter
\providecommand \@ifxundefined [1]{%
 \@ifx{#1\undefined}
}%
\providecommand \@ifnum [1]{%
 \ifnum #1\expandafter \@firstoftwo
 \else \expandafter \@secondoftwo
 \fi
}%
\providecommand \@ifx [1]{%
 \ifx #1\expandafter \@firstoftwo
 \else \expandafter \@secondoftwo
 \fi
}%
\providecommand \natexlab [1]{#1}%
\providecommand \enquote  [1]{``#1''}%
\providecommand \bibnamefont  [1]{#1}%
\providecommand \bibfnamefont [1]{#1}%
\providecommand \citenamefont [1]{#1}%
\providecommand \href@noop [0]{\@secondoftwo}%
\providecommand \href [0]{\begingroup \@sanitize@url \@href}%
\providecommand \@href[1]{\@@startlink{#1}\@@href}%
\providecommand \@@href[1]{\endgroup#1\@@endlink}%
\providecommand \@sanitize@url [0]{\catcode `\\12\catcode `\$12\catcode
  `\&12\catcode `\#12\catcode `\^12\catcode `\_12\catcode `\%12\relax}%
\providecommand \@@startlink[1]{}%
\providecommand \@@endlink[0]{}%
\providecommand \url  [0]{\begingroup\@sanitize@url \@url }%
\providecommand \@url [1]{\endgroup\@href {#1}{\urlprefix }}%
\providecommand \urlprefix  [0]{URL }%
\providecommand \Eprint [0]{\href }%
\providecommand \doibase [0]{https://doi.org/}%
\providecommand \selectlanguage [0]{\@gobble}%
\providecommand \bibinfo  [0]{\@secondoftwo}%
\providecommand \bibfield  [0]{\@secondoftwo}%
\providecommand \translation [1]{[#1]}%
\providecommand \BibitemOpen [0]{}%
\providecommand \bibitemStop [0]{}%
\providecommand \bibitemNoStop [0]{.\EOS\space}%
\providecommand \EOS [0]{\spacefactor3000\relax}%
\providecommand \BibitemShut  [1]{\csname bibitem#1\endcsname}%
\let\auto@bib@innerbib\@empty
%</preamble>
\bibitem [{\citenamefont {Schultz}\ and\ \citenamefont
  {Joachims}(2004)}]{schultz2004learning}%
  \BibitemOpen
  \bibfield  {author} {\bibinfo {author} {\bibfnamefont {M.}~\bibnamefont
  {Schultz}}\ and\ \bibinfo {author} {\bibfnamefont {T.}~\bibnamefont
  {Joachims}},\ }\bibfield  {title} {\bibinfo {title} {Learning a distance
  metric from relative comparisons},\ }\href@noop {} {\bibfield  {journal}
  {\bibinfo  {journal} {Advances in neural information processing systems}\
  }\textbf {\bibinfo {volume} {16}},\ \bibinfo {pages} {41} (\bibinfo {year}
  {2004})}\BibitemShut {NoStop}%
\bibitem [{\citenamefont {Shalev-Shwartz}\ \emph {et~al.}(2004)\citenamefont
  {Shalev-Shwartz}, \citenamefont {Singer},\ and\ \citenamefont
  {Ng}}]{shalev2004online}%
  \BibitemOpen
  \bibfield  {author} {\bibinfo {author} {\bibfnamefont {S.}~\bibnamefont
  {Shalev-Shwartz}}, \bibinfo {author} {\bibfnamefont {Y.}~\bibnamefont
  {Singer}},\ and\ \bibinfo {author} {\bibfnamefont {A.~Y.}\ \bibnamefont
  {Ng}},\ }\bibfield  {title} {\bibinfo {title} {Online and batch learning of
  pseudo-metrics},\ }in\ \href@noop {} {\emph {\bibinfo {booktitle}
  {Proceedings of the twenty-first international conference on Machine
  learning}}}\ (\bibinfo {year} {2004})\ p.~\bibinfo {pages} {94}\BibitemShut
  {NoStop}%
\bibitem [{\citenamefont {Chechik}\ \emph {et~al.}(2009)\citenamefont
  {Chechik}, \citenamefont {Sharma}, \citenamefont {Shalit},\ and\
  \citenamefont {Bengio}}]{chechik2009online}%
  \BibitemOpen
  \bibfield  {author} {\bibinfo {author} {\bibfnamefont {G.}~\bibnamefont
  {Chechik}}, \bibinfo {author} {\bibfnamefont {V.}~\bibnamefont {Sharma}},
  \bibinfo {author} {\bibfnamefont {U.}~\bibnamefont {Shalit}},\ and\ \bibinfo
  {author} {\bibfnamefont {S.}~\bibnamefont {Bengio}},\ }\bibfield  {title}
  {\bibinfo {title} {An online algorithm for large scale image similarity
  learning},\ }\href@noop {} {\  (\bibinfo {year} {2009})}\BibitemShut
  {NoStop}%
\bibitem [{\citenamefont {Bellet}\ \emph {et~al.}(2013)\citenamefont {Bellet},
  \citenamefont {Habrard},\ and\ \citenamefont {Sebban}}]{bellet2013survey}%
  \BibitemOpen
  \bibfield  {author} {\bibinfo {author} {\bibfnamefont {A.}~\bibnamefont
  {Bellet}}, \bibinfo {author} {\bibfnamefont {A.}~\bibnamefont {Habrard}},\
  and\ \bibinfo {author} {\bibfnamefont {M.}~\bibnamefont {Sebban}},\
  }\bibfield  {title} {\bibinfo {title} {A survey on metric learning for
  feature vectors and structured data},\ }\href@noop {} {\bibfield  {journal}
  {\bibinfo  {journal} {arXiv preprint arXiv:1306.6709}\ } (\bibinfo {year}
  {2013})}\BibitemShut {NoStop}%
\bibitem [{\citenamefont {Nicolae}\ \emph {et~al.}(2015)\citenamefont
  {Nicolae}, \citenamefont {Gaussier}, \citenamefont {Habrard},\ and\
  \citenamefont {Sebban}}]{nicolae2015joint}%
  \BibitemOpen
  \bibfield  {author} {\bibinfo {author} {\bibfnamefont {M.-I.}\ \bibnamefont
  {Nicolae}}, \bibinfo {author} {\bibfnamefont {{\'E}.}~\bibnamefont
  {Gaussier}}, \bibinfo {author} {\bibfnamefont {A.}~\bibnamefont {Habrard}},\
  and\ \bibinfo {author} {\bibfnamefont {M.}~\bibnamefont {Sebban}},\
  }\bibfield  {title} {\bibinfo {title} {Joint semi-supervised similarity
  learning for linear classification},\ }in\ \href@noop {} {\emph {\bibinfo
  {booktitle} {Joint European Conference on Machine Learning and Knowledge
  Discovery in Databases}}}\ (\bibinfo {organization} {Springer},\ \bibinfo
  {year} {2015})\ pp.\ \bibinfo {pages} {594--609}\BibitemShut {NoStop}%
\bibitem [{\citenamefont {Bengio}\ \emph {et~al.}(2003)\citenamefont {Bengio},
  \citenamefont {Paiement}, \citenamefont {Vincent}, \citenamefont {Delalleau},
  \citenamefont {Roux},\ and\ \citenamefont {Ouimet}}]{bengio2003out}%
  \BibitemOpen
  \bibfield  {author} {\bibinfo {author} {\bibfnamefont {Y.}~\bibnamefont
  {Bengio}}, \bibinfo {author} {\bibfnamefont {J.-f.}\ \bibnamefont
  {Paiement}}, \bibinfo {author} {\bibfnamefont {P.}~\bibnamefont {Vincent}},
  \bibinfo {author} {\bibfnamefont {O.}~\bibnamefont {Delalleau}}, \bibinfo
  {author} {\bibfnamefont {N.}~\bibnamefont {Roux}},\ and\ \bibinfo {author}
  {\bibfnamefont {M.}~\bibnamefont {Ouimet}},\ }\bibfield  {title} {\bibinfo
  {title} {Out-of-sample extensions for lle, isomap, mds, eigenmaps, and
  spectral clustering},\ }\href@noop {} {\bibfield  {journal} {\bibinfo
  {journal} {Advances in neural information processing systems}\ }\textbf
  {\bibinfo {volume} {16}},\ \bibinfo {pages} {177} (\bibinfo {year}
  {2003})}\BibitemShut {NoStop}%
\bibitem [{\citenamefont {Roweis}\ and\ \citenamefont
  {Saul}(2000)}]{roweis2000nonlinear}%
  \BibitemOpen
  \bibfield  {author} {\bibinfo {author} {\bibfnamefont {S.~T.}\ \bibnamefont
  {Roweis}}\ and\ \bibinfo {author} {\bibfnamefont {L.~K.}\ \bibnamefont
  {Saul}},\ }\bibfield  {title} {\bibinfo {title} {Nonlinear dimensionality
  reduction by locally linear embedding},\ }\href@noop {} {\bibfield  {journal}
  {\bibinfo  {journal} {science}\ }\textbf {\bibinfo {volume} {290}},\ \bibinfo
  {pages} {2323} (\bibinfo {year} {2000})}\BibitemShut {NoStop}%
\bibitem [{\citenamefont {Cox}\ and\ \citenamefont
  {Cox}(2008)}]{cox2008multidimensional}%
  \BibitemOpen
  \bibfield  {author} {\bibinfo {author} {\bibfnamefont {M.~A.}\ \bibnamefont
  {Cox}}\ and\ \bibinfo {author} {\bibfnamefont {T.~F.}\ \bibnamefont {Cox}},\
  }\bibfield  {title} {\bibinfo {title} {Multidimensional scaling},\ }in\
  \href@noop {} {\emph {\bibinfo {booktitle} {Handbook of data
  visualization}}}\ (\bibinfo  {publisher} {Springer},\ \bibinfo {year}
  {2008})\ pp.\ \bibinfo {pages} {315--347}\BibitemShut {NoStop}%
\bibitem [{\citenamefont {Belkin}\ and\ \citenamefont
  {Niyogi}(2003)}]{belkin2003laplacian}%
  \BibitemOpen
  \bibfield  {author} {\bibinfo {author} {\bibfnamefont {M.}~\bibnamefont
  {Belkin}}\ and\ \bibinfo {author} {\bibfnamefont {P.}~\bibnamefont
  {Niyogi}},\ }\bibfield  {title} {\bibinfo {title} {Laplacian eigenmaps for
  dimensionality reduction and data representation},\ }\href@noop {} {\bibfield
   {journal} {\bibinfo  {journal} {Neural computation}\ }\textbf {\bibinfo
  {volume} {15}},\ \bibinfo {pages} {1373} (\bibinfo {year}
  {2003})}\BibitemShut {NoStop}%
\bibitem [{\citenamefont {Kerenidis}\ \emph {et~al.}(2019)\citenamefont
  {Kerenidis}, \citenamefont {Landman},\ and\ \citenamefont
  {Prakash}}]{kerenidis2019quantum}%
  \BibitemOpen
  \bibfield  {author} {\bibinfo {author} {\bibfnamefont {I.}~\bibnamefont
  {Kerenidis}}, \bibinfo {author} {\bibfnamefont {J.}~\bibnamefont {Landman}},\
  and\ \bibinfo {author} {\bibfnamefont {A.}~\bibnamefont {Prakash}},\
  }\bibfield  {title} {\bibinfo {title} {Quantum algorithms for deep
  convolutional neural networks},\ }\href@noop {} {\bibfield  {journal}
  {\bibinfo  {journal} {arXiv preprint arXiv:1911.01117}\ } (\bibinfo {year}
  {2019})}\BibitemShut {NoStop}%
\bibitem [{\citenamefont {Dallaire-Demers}\ and\ \citenamefont
  {Killoran}(2018)}]{dallaire2018quantum}%
  \BibitemOpen
  \bibfield  {author} {\bibinfo {author} {\bibfnamefont {P.-L.}\ \bibnamefont
  {Dallaire-Demers}}\ and\ \bibinfo {author} {\bibfnamefont {N.}~\bibnamefont
  {Killoran}},\ }\bibfield  {title} {\bibinfo {title} {Quantum generative
  adversarial networks},\ }\href@noop {} {\bibfield  {journal} {\bibinfo
  {journal} {Physical Review A}\ }\textbf {\bibinfo {volume} {98}},\ \bibinfo
  {pages} {012324} (\bibinfo {year} {2018})}\BibitemShut {NoStop}%
\bibitem [{\citenamefont {Radha}(2021{\natexlab{a}})}]{radha2021quantum}%
  \BibitemOpen
  \bibfield  {author} {\bibinfo {author} {\bibfnamefont {S.~K.}\ \bibnamefont
  {Radha}},\ }\bibfield  {title} {\bibinfo {title} {Quantum constraint learning
  for quantum approximate optimization algorithm},\ }\href@noop {} {\bibfield
  {journal} {\bibinfo  {journal} {arXiv preprint arXiv:2105.06770}\ } (\bibinfo
  {year} {2021}{\natexlab{a}})}\BibitemShut {NoStop}%
\bibitem [{\citenamefont {Coyle}\ \emph {et~al.}(2021)\citenamefont {Coyle},
  \citenamefont {Henderson}, \citenamefont {Le}, \citenamefont {Kumar},
  \citenamefont {Paini},\ and\ \citenamefont {Kashefi}}]{coyle2021quantum}%
  \BibitemOpen
  \bibfield  {author} {\bibinfo {author} {\bibfnamefont {B.}~\bibnamefont
  {Coyle}}, \bibinfo {author} {\bibfnamefont {M.}~\bibnamefont {Henderson}},
  \bibinfo {author} {\bibfnamefont {J.~C.~J.}\ \bibnamefont {Le}}, \bibinfo
  {author} {\bibfnamefont {N.}~\bibnamefont {Kumar}}, \bibinfo {author}
  {\bibfnamefont {M.}~\bibnamefont {Paini}},\ and\ \bibinfo {author}
  {\bibfnamefont {E.}~\bibnamefont {Kashefi}},\ }\bibfield  {title} {\bibinfo
  {title} {Quantum versus classical generative modelling in finance},\
  }\href@noop {} {\bibfield  {journal} {\bibinfo  {journal} {Quantum Science
  and Technology}\ }\textbf {\bibinfo {volume} {6}},\ \bibinfo {pages} {024013}
  (\bibinfo {year} {2021})}\BibitemShut {NoStop}%
\bibitem [{\citenamefont {Radha}(2021{\natexlab{b}})}]{radha2021quantumwick}%
  \BibitemOpen
  \bibfield  {author} {\bibinfo {author} {\bibfnamefont {S.~K.}\ \bibnamefont
  {Radha}},\ }\bibfield  {title} {\bibinfo {title} {Quantum option pricing
  using wick rotated imaginary time evolution},\ }\href@noop {} {\bibfield
  {journal} {\bibinfo  {journal} {arXiv preprint arXiv:2101.04280}\ } (\bibinfo
  {year} {2021}{\natexlab{b}})}\BibitemShut {NoStop}%
\bibitem [{\citenamefont {Liu}\ \emph {et~al.}(2021)\citenamefont {Liu},
  \citenamefont {Arunachalam},\ and\ \citenamefont {Temme}}]{liu2021rigorous}%
  \BibitemOpen
  \bibfield  {author} {\bibinfo {author} {\bibfnamefont {Y.}~\bibnamefont
  {Liu}}, \bibinfo {author} {\bibfnamefont {S.}~\bibnamefont {Arunachalam}},\
  and\ \bibinfo {author} {\bibfnamefont {K.}~\bibnamefont {Temme}},\ }\bibfield
   {title} {\bibinfo {title} {A rigorous and robust quantum speed-up in
  supervised machine learning},\ }\href@noop {} {\bibfield  {journal} {\bibinfo
   {journal} {Nature Physics}\ }\textbf {\bibinfo {volume} {17}},\ \bibinfo
  {pages} {1013} (\bibinfo {year} {2021})}\BibitemShut {NoStop}%
\bibitem [{\citenamefont {Sweke}\ \emph {et~al.}(2021)\citenamefont {Sweke},
  \citenamefont {Seifert}, \citenamefont {Hangleiter},\ and\ \citenamefont
  {Eisert}}]{sweke2021quantum}%
  \BibitemOpen
  \bibfield  {author} {\bibinfo {author} {\bibfnamefont {R.}~\bibnamefont
  {Sweke}}, \bibinfo {author} {\bibfnamefont {J.-P.}\ \bibnamefont {Seifert}},
  \bibinfo {author} {\bibfnamefont {D.}~\bibnamefont {Hangleiter}},\ and\
  \bibinfo {author} {\bibfnamefont {J.}~\bibnamefont {Eisert}},\ }\bibfield
  {title} {\bibinfo {title} {On the quantum versus classical learnability of
  discrete distributions},\ }\href@noop {} {\bibfield  {journal} {\bibinfo
  {journal} {Quantum}\ }\textbf {\bibinfo {volume} {5}},\ \bibinfo {pages}
  {417} (\bibinfo {year} {2021})}\BibitemShut {NoStop}%
\bibitem [{\citenamefont {Huang}\ \emph {et~al.}(2021)\citenamefont {Huang},
  \citenamefont {Broughton}, \citenamefont {Mohseni}, \citenamefont {Babbush},
  \citenamefont {Boixo}, \citenamefont {Neven},\ and\ \citenamefont
  {McClean}}]{huang2021power}%
  \BibitemOpen
  \bibfield  {author} {\bibinfo {author} {\bibfnamefont {H.-Y.}\ \bibnamefont
  {Huang}}, \bibinfo {author} {\bibfnamefont {M.}~\bibnamefont {Broughton}},
  \bibinfo {author} {\bibfnamefont {M.}~\bibnamefont {Mohseni}}, \bibinfo
  {author} {\bibfnamefont {R.}~\bibnamefont {Babbush}}, \bibinfo {author}
  {\bibfnamefont {S.}~\bibnamefont {Boixo}}, \bibinfo {author} {\bibfnamefont
  {H.}~\bibnamefont {Neven}},\ and\ \bibinfo {author} {\bibfnamefont {J.~R.}\
  \bibnamefont {McClean}},\ }\bibfield  {title} {\bibinfo {title} {Power of
  data in quantum machine learning},\ }\href@noop {} {\bibfield  {journal}
  {\bibinfo  {journal} {Nature communications}\ }\textbf {\bibinfo {volume}
  {12}},\ \bibinfo {pages} {1} (\bibinfo {year} {2021})}\BibitemShut {NoStop}%
\bibitem [{\citenamefont {Schuld}(2021)}]{schuld2021quantum}%
  \BibitemOpen
  \bibfield  {author} {\bibinfo {author} {\bibfnamefont {M.}~\bibnamefont
  {Schuld}},\ }\bibfield  {title} {\bibinfo {title} {Quantum machine learning
  models are kernel methods},\ }\href@noop {} {\bibfield  {journal} {\bibinfo
  {journal} {arXiv e-prints}\ ,\ \bibinfo {pages} {arXiv}} (\bibinfo {year}
  {2021})}\BibitemShut {NoStop}%
\bibitem [{\citenamefont {Pronobis}\ and\ \citenamefont
  {M{\"u}ller}(2020)}]{pronobis2020kernel}%
  \BibitemOpen
  \bibfield  {author} {\bibinfo {author} {\bibfnamefont {W.}~\bibnamefont
  {Pronobis}}\ and\ \bibinfo {author} {\bibfnamefont {K.-R.}\ \bibnamefont
  {M{\"u}ller}},\ }\bibfield  {title} {\bibinfo {title} {Kernel methods for
  quantum chemistry},\ }in\ \href@noop {} {\emph {\bibinfo {booktitle} {Machine
  Learning Meets Quantum Physics}}}\ (\bibinfo  {publisher} {Springer},\
  \bibinfo {year} {2020})\ pp.\ \bibinfo {pages} {25--36}\BibitemShut {NoStop}%
\bibitem [{\citenamefont {Lloyd}\ \emph
  {et~al.}(2020{\natexlab{a}})\citenamefont {Lloyd}, \citenamefont {Schuld},
  \citenamefont {Ijaz}, \citenamefont {Izaac},\ and\ \citenamefont
  {Killoran}}]{lloyd2020quantum}%
  \BibitemOpen
  \bibfield  {author} {\bibinfo {author} {\bibfnamefont {S.}~\bibnamefont
  {Lloyd}}, \bibinfo {author} {\bibfnamefont {M.}~\bibnamefont {Schuld}},
  \bibinfo {author} {\bibfnamefont {A.}~\bibnamefont {Ijaz}}, \bibinfo {author}
  {\bibfnamefont {J.}~\bibnamefont {Izaac}},\ and\ \bibinfo {author}
  {\bibfnamefont {N.}~\bibnamefont {Killoran}},\ }\href@noop {} {\bibinfo
  {title} {Quantum embeddings for machine learning}} (\bibinfo {year}
  {2020}{\natexlab{a}}),\ \Eprint {https://arxiv.org/abs/2001.03622}
  {arXiv:2001.03622 [quant-ph]} \BibitemShut {NoStop}%
\bibitem [{\citenamefont {Bromley}\ \emph {et~al.}(1993)\citenamefont
  {Bromley}, \citenamefont {Bentz}, \citenamefont {Bottou}, \citenamefont
  {Guyon}, \citenamefont {LeCun}, \citenamefont {Moore}, \citenamefont
  {S{\"a}ckinger},\ and\ \citenamefont {Shah}}]{bromley1993signature}%
  \BibitemOpen
  \bibfield  {author} {\bibinfo {author} {\bibfnamefont {J.}~\bibnamefont
  {Bromley}}, \bibinfo {author} {\bibfnamefont {J.~W.}\ \bibnamefont {Bentz}},
  \bibinfo {author} {\bibfnamefont {L.}~\bibnamefont {Bottou}}, \bibinfo
  {author} {\bibfnamefont {I.}~\bibnamefont {Guyon}}, \bibinfo {author}
  {\bibfnamefont {Y.}~\bibnamefont {LeCun}}, \bibinfo {author} {\bibfnamefont
  {C.}~\bibnamefont {Moore}}, \bibinfo {author} {\bibfnamefont
  {E.}~\bibnamefont {S{\"a}ckinger}},\ and\ \bibinfo {author} {\bibfnamefont
  {R.}~\bibnamefont {Shah}},\ }\bibfield  {title} {\bibinfo {title} {Signature
  verification using a “siamese” time delay neural network},\ }\href@noop
  {} {\bibfield  {journal} {\bibinfo  {journal} {International Journal of
  Pattern Recognition and Artificial Intelligence}\ }\textbf {\bibinfo {volume}
  {7}},\ \bibinfo {pages} {669} (\bibinfo {year} {1993})}\BibitemShut {NoStop}%
\bibitem [{\citenamefont {Havl{\'\i}{\v{c}}ek}\ \emph
  {et~al.}(2019)\citenamefont {Havl{\'\i}{\v{c}}ek}, \citenamefont
  {C{\'o}rcoles}, \citenamefont {Temme}, \citenamefont {Harrow}, \citenamefont
  {Kandala}, \citenamefont {Chow},\ and\ \citenamefont
  {Gambetta}}]{havlivcek2019supervised}%
  \BibitemOpen
  \bibfield  {author} {\bibinfo {author} {\bibfnamefont {V.}~\bibnamefont
  {Havl{\'\i}{\v{c}}ek}}, \bibinfo {author} {\bibfnamefont {A.~D.}\
  \bibnamefont {C{\'o}rcoles}}, \bibinfo {author} {\bibfnamefont
  {K.}~\bibnamefont {Temme}}, \bibinfo {author} {\bibfnamefont {A.~W.}\
  \bibnamefont {Harrow}}, \bibinfo {author} {\bibfnamefont {A.}~\bibnamefont
  {Kandala}}, \bibinfo {author} {\bibfnamefont {J.~M.}\ \bibnamefont {Chow}},\
  and\ \bibinfo {author} {\bibfnamefont {J.~M.}\ \bibnamefont {Gambetta}},\
  }\bibfield  {title} {\bibinfo {title} {Supervised learning with
  quantum-enhanced feature spaces},\ }\href@noop {} {\bibfield  {journal}
  {\bibinfo  {journal} {Nature}\ }\textbf {\bibinfo {volume} {567}},\ \bibinfo
  {pages} {209} (\bibinfo {year} {2019})}\BibitemShut {NoStop}%
\bibitem [{\citenamefont {Balcan}\ \emph {et~al.}(2008)\citenamefont {Balcan},
  \citenamefont {Blum},\ and\ \citenamefont {Srebro}}]{balcan2008improved}%
  \BibitemOpen
  \bibfield  {author} {\bibinfo {author} {\bibfnamefont {M.-F.}\ \bibnamefont
  {Balcan}}, \bibinfo {author} {\bibfnamefont {A.}~\bibnamefont {Blum}},\ and\
  \bibinfo {author} {\bibfnamefont {N.}~\bibnamefont {Srebro}},\ }\bibfield
  {title} {\bibinfo {title} {Improved guarantees for learning via similarity
  functions},\ }\href@noop {} {\  (\bibinfo {year} {2008})}\BibitemShut
  {NoStop}%
\bibitem [{\citenamefont {Lloyd}\ \emph
  {et~al.}(2020{\natexlab{b}})\citenamefont {Lloyd}, \citenamefont {Schuld},
  \citenamefont {Ijaz}, \citenamefont {Izaac},\ and\ \citenamefont
  {Killoran}}]{Lloyd2020}%
  \BibitemOpen
  \bibfield  {author} {\bibinfo {author} {\bibfnamefont {S.}~\bibnamefont
  {Lloyd}}, \bibinfo {author} {\bibfnamefont {M.}~\bibnamefont {Schuld}},
  \bibinfo {author} {\bibfnamefont {A.}~\bibnamefont {Ijaz}}, \bibinfo {author}
  {\bibfnamefont {J.}~\bibnamefont {Izaac}},\ and\ \bibinfo {author}
  {\bibfnamefont {N.}~\bibnamefont {Killoran}},\ }\bibfield  {title} {\bibinfo
  {title} {Quantum embeddings for machine learning},\ }\href@noop {} {\
  (\bibinfo {year} {2020}{\natexlab{b}})},\ \Eprint
  {https://arxiv.org/abs/2001.03622} {arXiv:2001.03622 [quant-ph]} \BibitemShut
  {NoStop}%
\bibitem [{\citenamefont {Hubregtsen}\ \emph {et~al.}(2021)\citenamefont
  {Hubregtsen}, \citenamefont {Wierichs}, \citenamefont {Gil-Fuster},
  \citenamefont {Derks}, \citenamefont {Faehrmann},\ and\ \citenamefont
  {Meyer}}]{Hubregtsen2021}%
  \BibitemOpen
  \bibfield  {author} {\bibinfo {author} {\bibfnamefont {T.}~\bibnamefont
  {Hubregtsen}}, \bibinfo {author} {\bibfnamefont {D.}~\bibnamefont
  {Wierichs}}, \bibinfo {author} {\bibfnamefont {E.}~\bibnamefont
  {Gil-Fuster}}, \bibinfo {author} {\bibfnamefont {P.-J. H.~S.}\ \bibnamefont
  {Derks}}, \bibinfo {author} {\bibfnamefont {P.~K.}\ \bibnamefont
  {Faehrmann}},\ and\ \bibinfo {author} {\bibfnamefont {J.~J.}\ \bibnamefont
  {Meyer}},\ }\bibfield  {title} {\bibinfo {title} {Training quantum embedding
  kernels on near-term quantum computers},\ }\href@noop {} {\  (\bibinfo {year}
  {2021})},\ \Eprint {https://arxiv.org/abs/2105.02276} {arXiv:2105.02276
  [quant-ph]} \BibitemShut {NoStop}%
\bibitem [{\citenamefont {Lloyd}\ \emph {et~al.}(2013)\citenamefont {Lloyd},
  \citenamefont {Mohseni},\ and\ \citenamefont
  {Rebentrost}}]{lloyd2013quantum}%
  \BibitemOpen
  \bibfield  {author} {\bibinfo {author} {\bibfnamefont {S.}~\bibnamefont
  {Lloyd}}, \bibinfo {author} {\bibfnamefont {M.}~\bibnamefont {Mohseni}},\
  and\ \bibinfo {author} {\bibfnamefont {P.}~\bibnamefont {Rebentrost}},\
  }\href@noop {} {\bibinfo {title} {Quantum algorithms for supervised and
  unsupervised machine learning}} (\bibinfo {year} {2013}),\ \Eprint
  {https://arxiv.org/abs/1307.0411} {arXiv:1307.0411 [quant-ph]} \BibitemShut
  {NoStop}%
\bibitem [{\citenamefont {Powell}(2007)}]{powell2007view}%
  \BibitemOpen
  \bibfield  {author} {\bibinfo {author} {\bibfnamefont {M.~J.}\ \bibnamefont
  {Powell}},\ }\bibfield  {title} {\bibinfo {title} {A view of algorithms for
  optimization without derivatives},\ }\href@noop {} {\bibfield  {journal}
  {\bibinfo  {journal} {Mathematics Today-Bulletin of the Institute of
  Mathematics and its Applications}\ }\textbf {\bibinfo {volume} {43}},\
  \bibinfo {pages} {170} (\bibinfo {year} {2007})}\BibitemShut {NoStop}%
\bibitem [{\citenamefont {Shiina}\ \emph {et~al.}(2020)\citenamefont {Shiina},
  \citenamefont {Mori}, \citenamefont {Okabe},\ and\ \citenamefont
  {Lee}}]{shiina2020machine}%
  \BibitemOpen
  \bibfield  {author} {\bibinfo {author} {\bibfnamefont {K.}~\bibnamefont
  {Shiina}}, \bibinfo {author} {\bibfnamefont {H.}~\bibnamefont {Mori}},
  \bibinfo {author} {\bibfnamefont {Y.}~\bibnamefont {Okabe}},\ and\ \bibinfo
  {author} {\bibfnamefont {H.~K.}\ \bibnamefont {Lee}},\ }\bibfield  {title}
  {\bibinfo {title} {Machine-learning studies on spin models},\ }\href@noop {}
  {\bibfield  {journal} {\bibinfo  {journal} {Scientific reports}\ }\textbf
  {\bibinfo {volume} {10}},\ \bibinfo {pages} {1} (\bibinfo {year}
  {2020})}\BibitemShut {NoStop}%
\bibitem [{\citenamefont {Bai}\ \emph {et~al.}(2019)\citenamefont {Bai},
  \citenamefont {Rossi}, \citenamefont {Cui}, \citenamefont {Cheng},\ and\
  \citenamefont {Hancock}}]{bai2019quantum}%
  \BibitemOpen
  \bibfield  {author} {\bibinfo {author} {\bibfnamefont {L.}~\bibnamefont
  {Bai}}, \bibinfo {author} {\bibfnamefont {L.}~\bibnamefont {Rossi}}, \bibinfo
  {author} {\bibfnamefont {L.}~\bibnamefont {Cui}}, \bibinfo {author}
  {\bibfnamefont {J.}~\bibnamefont {Cheng}},\ and\ \bibinfo {author}
  {\bibfnamefont {E.~R.}\ \bibnamefont {Hancock}},\ }\bibfield  {title}
  {\bibinfo {title} {A quantum-inspired similarity measure for the analysis of
  complete weighted graphs},\ }\href@noop {} {\bibfield  {journal} {\bibinfo
  {journal} {IEEE transactions on cybernetics}\ }\textbf {\bibinfo {volume}
  {50}},\ \bibinfo {pages} {1264} (\bibinfo {year} {2019})}\BibitemShut
  {NoStop}%
\bibitem [{\citenamefont {Benedetti}\ \emph {et~al.}(2019)\citenamefont
  {Benedetti}, \citenamefont {Lloyd}, \citenamefont {Sack},\ and\ \citenamefont
  {Fiorentini}}]{benedetti2019parameterized}%
  \BibitemOpen
  \bibfield  {author} {\bibinfo {author} {\bibfnamefont {M.}~\bibnamefont
  {Benedetti}}, \bibinfo {author} {\bibfnamefont {E.}~\bibnamefont {Lloyd}},
  \bibinfo {author} {\bibfnamefont {S.}~\bibnamefont {Sack}},\ and\ \bibinfo
  {author} {\bibfnamefont {M.}~\bibnamefont {Fiorentini}},\ }\bibfield  {title}
  {\bibinfo {title} {Parameterized quantum circuits as machine learning
  models},\ }\href@noop {} {\bibfield  {journal} {\bibinfo  {journal} {Quantum
  Science and Technology}\ }\textbf {\bibinfo {volume} {4}},\ \bibinfo {pages}
  {043001} (\bibinfo {year} {2019})}\BibitemShut {NoStop}%
\bibitem [{\citenamefont {Bottou}(2010)}]{bottou2010large}%
  \BibitemOpen
  \bibfield  {author} {\bibinfo {author} {\bibfnamefont {L.}~\bibnamefont
  {Bottou}},\ }\bibfield  {title} {\bibinfo {title} {Large-scale machine
  learning with stochastic gradient descent},\ }in\ \href@noop {} {\emph
  {\bibinfo {booktitle} {Proceedings of COMPSTAT'2010}}}\ (\bibinfo
  {publisher} {Springer},\ \bibinfo {year} {2010})\ pp.\ \bibinfo {pages}
  {177--186}\BibitemShut {NoStop}%
\end{thebibliography}%


%apsrev4-2.bst 2019-01-14 (MD) hand-edited version of apsrev4-1.bst
%Control: key (0)
%Control: author (8) initials jnrlst
%Control: editor formatted (1) identically to author
%Control: production of article title (0) allowed
%Control: page (0) single
%Control: year (1) truncated
%Control: production of eprint (0) enabled
%
\end{document}